\documentclass{article}

\usepackage{amsmath, amsfonts}
\usepackage[table]{xcolor}
\usepackage[a4paper, total={7in, 10in}]{geometry}
\usepackage{ulem}
\usepackage{graphicx}
\usepackage{mathtools}
\usepackage{siunitx}
\usepackage{wrapfig}
\usepackage{algpseudocode}
\usepackage{algorithm}
\usepackage{bm}
\usepackage{comment}
\usepackage{hyperref}

\hypersetup{
    colorlinks=true,
    linkcolor=blue,
    filecolor=magenta,      
    urlcolor=cyan,
    pdfpagemode=FullScreen,
}

\begin{document}

\title{Numerical Identification of Stationary States and Their Stability in a Model of Quantum Droplets}
\author{
  Sun Lee\thanks{Department of Mathematics, Pennsylvania State University, University Park, PA, USA. Email: skl5876@psu.edu} 
  \and 
  Panayotis G. Kevrekidis\thanks{Department of Mathematics and Statistics, University of Massachusetts Amherst, MA, USA. Email: kevrekid@umass.edu} 
  \thanks{Department of Physics, University of Massachusetts Amherst, MA, USA}
  \thanks{Theoretical Sciences Visiting Program, Okinawa Institute of Science and Technology Graduate University, Onna, 904-0495, Japan}
  \and 
  Wenrui Hao\thanks{Department of Mathematics, Pennsylvania State University, University Park, PA, USA. Email: wxh64@psu.edu}
}

\date{}

\maketitle

\begin{abstract}
    In this work, we are motivated by a recent variant of the nonlinear Schrödinger (NLS) equation describing cold, dilute atomic condensates with quantum fluctuation effects. Our goal is to develop robust numerical methods capable of uncovering diverse stationary solutions in such NLS models. Specifically, and in line with recent theoretical and experimental interest, we focus on ultracold quantum droplets in Bose mixtures influenced by the Lee–Huang–Yang quantum fluctuation correction and study these systems in one- and two-dimensional settings. To this end, we deploy several numerical techniques. The homotopy grid method allows systematic refinement from coarse to fine spatial discretizations in one dimension, while the dimension-by-dimension homotopy approach extends one-dimensional solutions to two-dimensional domains. These methods effectively detect broad families of stationary states, many of which have not been previously reported, to the best of our knowledge. Furthermore, they enable the monitoring of solution continuation and bifurcation phenomena. During our investigation, we encounter unusual bifurcation events, including nonstandard pitchforks and saddle-center bifurcations, which exhibit novel stability transitions. For example, we identify continuous pathways connecting vortex and dark soliton stripe branches, absent in the standard cubic defocusing model. Overall, the presence of competing mean-field and quantum fluctuation interactions leads to a richer bifurcation structure than in traditional cubic NLS systems. These findings suggest that similar complex bifurcation and stability phenomena may appear in other settings, including higher-dimensional systems or models with competing nonlinearities such as cubic-quintic interactions, highlighting the importance of further theoretical and numerical exploration.

\end{abstract}



\section{Introduction}

The realm of ultracold atomic condensates has constituted over the past three decades a pristine platform enabling
the extensive tunability of external potentials and interatomic interactions~\cite{kevrekidis2015defocusing,pethick2008bose,PitaevskiiStringari2016}. Over the past few years, one of the contexts within such systems that has gained considerable traction has concerned the study of ultradilute self-bound states of matter, most notably in the form of quantum droplets~\cite{luo2021new,mistakidis2023few}. The experimental detection of such states has been realized not only in dipolar gases~\cite{bottcher2020new,chomaz2022dipolar}, but also in the principal setup of interest in the present work, namely in the realm of Bose mixtures~\cite{burchianti2020dual,cabrera2018quantum,cheiney2018bright,d2019observation,semeghini2018self}, where it continues to be a topic of active experimental investigation~\cite{cavicchioli2024dynamicalformationmultiplequantum}.

From a theoretical perspective, the emergence of such droplet states is due to the competition between attractive and repulsive interactions: these arise from the interplay of the mean-field-nonlinearity (due to interatomic interactions) and the role of quantum fluctuations described by the famous Lee-Huang-Yang (LHY) correction~\cite{larsen1963binary,lee1957eigenvalues}. A key feature of the latter is how its nature, both in terms of its functional form and even its sign, is affected by the effective dimensionality of the problem~\cite{Ilg_crossover_2018}. Accordingly, through the work of Petrov (among others)~~\cite{petrov2015quantum,petrov2016ultradilute}, it was recognized that this competition leads to an extended form of the well-established Gross-Pitaevskii equation, the so-called eGPE that will be the central model of the present investigation. It should be added here that within the latter model, numerous nonlinear excitations have been identified and their stability has been explored, most notably in one spatial dimension, including dark soliton and multi-soliton states~\cite{edmonds2023dark,katsimiga2023solitary}, the so-called (unstable) bubble states~\cite{katsimiga2023interactions}, the intriguing kink states~\cite{kartashov2022spinor,katsimiga2023interactions,tylutki2020collective} that have recently been argued to even be dynamically robust in higher dimensions~\cite{mistakidis2024generictransversestabilitykink}, as well as genuinely higher dimensional structures such as vortices~\cite{li2018two,yougurt2023vortex}. Many of the developments in this field (especially the early ones) have been summarized in the review work of~\cite{luo2021new}.

In the present work, we leverage this important and timely theoretical model—one that has received considerable experimental interest from the atomic Bose–Einstein condensate (BEC) community—to investigate the potential of several numerical methods for systematically characterizing the extensive variety of states and rich bifurcation structures inherent to this class of models. In one spatial dimension, we propose two methods that perform well for the purposes of our study. The first is a companion-based multi-level strategy, in which we double the number of grid points from level $l$ to level $l+1$; by solving for the newly introduced intermediate points, we obtain a high-quality initial guess for the next iteration level. This approach, inspired by the companion-based framework in \cite{hao2024companion}, offers a natural mechanism for constructing increasingly refined solution approximations by effectively utilizing information from coarser grids.

Another method is the so-called homotopy grid expansion method, which begins with two solutions computed on a coarser grid and combines them by tracking an appropriate homotopy problem as the homotopy parameter varies from $s=0$ to $s=1$, with the goal of obtaining a finer-grid solution of the original problem. The intuition is that by continuously deforming a known solution pair into a candidate finer-grid solution, one can overcome convergence difficulties that often occur when attempting to directly solve the refined system. This approach is connected to general homotopy continuation techniques for nonlinear equations and bifurcation problems~\cite{allgower2012numerical,seydel2009practical}, which have been successfully adapted to grid-refinement strategies in PDE settings.

Finally, in higher dimensions, we leverage solutions from the one-dimensional system and develop a dimension-by-dimension homotopy method, which allows us to obtain appropriately refined, full solutions to the original problem. This continuation proceeds in a manner analogous to the finer-grid refinement used in the homotopy grid expansion, but here the refinement occurs along previously absent dimensions, gradually extending the solution from lower- to higher-dimensional states~\cite{allgower2012numerical,seydel2009practical}.

Our presentation of these numerical methods and their results is organized as follows. In Section~\ref{sec:Model}, we provide a brief recap of the modeling setup to set the stage for our computations. In Section~\ref{sec:Numerical}, we give detailed descriptions of the aforementioned one-dimensional (the first two) and two-dimensional (the third) numerical methods. Our main findings are summarized in Section~\ref{sec:Results}, where we show how the proposed methods allow us to construct bifurcation diagrams closely resembling the current state-of-the-art results for the one-dimensional case~\cite{katsimiga2023solitary}. 

The one-dimensional case serves both as a validation study of the proposed numerical techniques and as a warm-up for the far more complex two-dimensional solution landscape. However, the results we obtain are not identical to previously known outcomes, as we focus on quasi-one-dimensional solutions (i.e., $y$-independent) of the fully two-dimensional problem. We then extend our considerations to constructing corresponding bifurcation diagrams in two-dimensional settings. 

In doing so, we observe that, while some solution branches are analogous to those known from the cubic case, the two-dimensional scenario exhibits a sequence of bifurcation phenomena with numerous unexpected and atypical features. We comment on these observations to the best of our understanding, highlighting both the anticipated similarities and the key differences relative to the more well-studied cubic, defocusing nonlinearity case~\cite{charalampidis2018computing,MIDDELKAMP20111449}.

We also find that this setting raises several intriguing questions for further study. We believe that the results presented in this paper establish the realm of competing nonlinearities in the presence of a trapping potential as an exceptionally rich and diverse platform for nonlinear waveforms, with the potential for near-future experimental realizations of this droplet-bearing system. Finally, in Section~5, we summarize our findings and present our conclusions, along with several directions for future research. The Appendix is reserved for a set of benchmark studies demonstrating that our methods are capable of reproducing previously reported results for the cubic Gross–Pitaevskii equation in the presence of a parabolic trap.

\section{Model Problem}\label{sec:Model}

\subsection{Problem Formulation}

We consider the dynamics of a single-component system that has been widely studied in the context of ultracold Bose mixtures~\cite{luo2021new} within a bounded, $n$-dimensional rectangular domain $D=(-d,d)^n \subset \mathbb{R}^n$. The state of the system is described by a complex-valued wave function, $\Psi(\bm{x},t): D \times \mathbb{R}^+ \rightarrow \mathbb{C}$, whose evolution is governed by a variant of the nonlinear Schrödinger equation (NLS):
\begin{equation}
    i \partial_t \Psi = -\Delta \Psi + \mathcal{N}(\Psi) + V(r) \Psi, \quad \mbox{in } D \times \mathbb{R}^+, \label{NLS}
\end{equation}
where $r = \|\bm{x}\|$. Here, $V(r)$ denotes an external potential—typically the parabolic quantum harmonic oscillator potential used throughout this paper, and $\mathcal{N}(\Psi)$ is a nonlinear term modeling the wave dynamics of the atomic system~\cite{kevrekidis2015defocusing,PitaevskiiStringari2016}.

The primary examples we consider include the extended Gross–Pitaevskii (eGPE) model with competing quadratic–cubic nonlinearity in 1D, $\mathcal{N}(\Psi)=-|\Psi|\Psi+ |\Psi|^2\Psi$, and a logarithmic nonlinearity in 2D, $\mathcal{N}(\Psi)=|\Psi|^2 \ln(|\Psi|^2)\Psi$~\cite{luo2021new}. The logarithmic nonlinearity in 2D captures the competition between attraction at low densities (recall that the density is defined as $|\Psi|^2$) and repulsion at higher densities, due to the change of sign in the logarithmic term. Additionally, we revisit the purely cubic nonlinearity, $\mathcal{N}(\Psi)=|\Psi|^2\Psi$, previously explored in 1D~\cite{alfimov2007nonlinear} and 2D~\cite{MIDDELKAMP20111449}, as a benchmark example in Appendix~\ref{sec:appendix_name}.

To find stationary solutions of Eq.~\eqref{NLS}, we employ the standard ansatz~\cite{alfimov2007nonlinear,charalampidis2018computing}:
\begin{equation}
    \Psi(\bm{x},t) = \psi(\bm{x}) \exp(-i \mu t), \label{ansatz}
\end{equation}
where $\psi(\bm{x})$ is the time-independent spatial profile, and $\mu$ is a real constant representing the chemical potential, treated as an eigenvalue of the nonlinear problem. Substituting this ansatz into Eq.~\eqref{NLS} yields the stationary equation:
\begin{equation}\label{Eq1}
    -\Delta \psi + V(r)\psi + \mathcal{N}(\psi) - \mu \psi = 0, \quad \mbox{in } D.
\end{equation}
We seek spatially localized solutions. For numerical approximation, we solve Eq.~\eqref{Eq1} on a sufficiently large domain $D$ and impose homogeneous Dirichlet boundary conditions on its boundary $\partial D$:
\begin{equation}\label{main:bdry} 
    \psi = 0, \quad \mbox{on } \partial D.
\end{equation}
Given the confining parabolic trap and vanishing density at large distances, these boundary conditions are appropriate.

For convenience, we define
\begin{equation}
    f(r, \psi) = V(r)\psi + \mathcal{N}(\psi) - \mu \psi,
\end{equation}
so that Eq.~\eqref{Eq1} can be written in the standard elliptic form:
\begin{equation}\label{Eq2}
    -\Delta \psi + f(r, \psi) = 0, \quad \mbox{in } D.
\end{equation}

For numerical computations, it is often advantageous to decompose the complex function $\psi$ into its real and imaginary parts, $\psi = \psi_{\mathbb{R}} + i\psi_{\mathbb{C}}$. This transforms the single complex PDE into a system of two coupled, real-valued nonlinear equations:
\begin{equation}\label{main}
\begin{cases}
    -\Delta \psi_{\mathbb{R}} + f_\mathbb{R}(r, \psi_{\mathbb{R}}, \psi_{\mathbb{C}}) = 0, \\
    -\Delta \psi_{\mathbb{C}} + f_\mathbb{C}(r, \psi_{\mathbb{R}}, \psi_{\mathbb{C}}) = 0, 
\end{cases}
\quad \mbox{in } D,
\end{equation}
where $f_\mathbb{R}$ and $f_\mathbb{C}$ denote the real and imaginary components of $f$, respectively.

\subsection{Stability}\label{stability}

 For the stability, our starting point is 
 a stationary solution $\psi_0$, which will be perturbed as
 follows~\cite{kevrekidis2015defocusing}:
  \begin{equation}\label{pert}
     \tilde{\psi}=e^{-i\mu t}(\psi_0+\epsilon(ae^{i\omega t}+b^*e^{-i\omega^* t})),
 \end{equation}
 in order to examine the fate of small perturbations controlled by the formal, small parameter $\epsilon$ over time.
Here, $^*$ denotes complex conjugation. We start with specific examples when $\mathcal{N}(\Psi)=|\Psi|^2 \ln(|\Psi|^2)\Psi $ as
\begin{equation}\label{ex_st}
     i \partial_t \Psi =-\frac{\triangle \Psi }{2} + |\Psi|^2\Psi \ln(|\Psi|^2) +V(r)\Psi , \quad \mbox{ in } D \times \mathbb{R}^+.
\end{equation}

Inserting \eqref{pert} into equation \eqref{ex_st} and substituting within the density expression, we obtain

    \begin{equation}|\tilde{\psi}|^2=\tilde{\psi}\tilde{\psi}^*=\psi_0\psi^*_0+\epsilon\left((a\psi_0^*+b\psi_0)e^{i\omega t}+(a^*\psi_0+b^*\psi_0^*)e^{-i\omega^*t}\right)+O(\epsilon^2)
 \end{equation}
From the Taylor expansion $\ln(x+\epsilon)=\ln(x)+\frac{\epsilon}{x}-\frac{\epsilon^2}{2x^2}+\cdots$, we can deduce
     \begin{equation}
   \ln(|\tilde{\psi}|^2)=\ln(\psi_0\psi_0^*)+\frac{\epsilon\left((a\psi_0^*+b\psi_0)e^{i\omega t}+(a^*\psi_0+b^*\psi_0^*)e^{-i\omega^*t}\right)}{\psi_0\psi_0^*}+O(\epsilon^2)
 \end{equation}

Through combining the relevant results and neglecting the
terms of $O(\epsilon^2)$, we obtain the 
linearization operators:

\begin{eqnarray}
    A_{11} &=& -\frac{\triangle }{2} + |\psi_0|^2(2\ln{(|\psi_0|^2)}+1)+V(r) - \mu 
\\
    A_{12} &=& \psi_0^2(\ln{(|\psi_0|^2)}+1)
\end{eqnarray}
which arise in the context of the resulting eigenvalue problem as:
\begin{equation}\label{final_mat}
-\omega \begin{pmatrix}
 a\\b
\end{pmatrix}=
    \begin{pmatrix}
  A_{11} & A_{12}\\-A^*_{12} & -A_{11}
\end{pmatrix}
\begin{pmatrix}
 a\\b
\end{pmatrix}.
\end{equation}
If the eigenfrequency $\omega$ is purely real then that implies that the solution is stable. On the other hand, a purely imaginary eigenvalue $\omega$ leads to exponential growth or decay. A complex eigenvalue with both real and imaginary parts gives rise to oscillatory instabilities, where exponential growth/decay is combined with oscillations.

\section{Numerical Methods}\label{sec:Numerical}

In this section, we present numerical methods for computing multiple solutions of Eq.~\eqref{main} in both one and two spatial dimensions. For the one-dimensional case, we employ a companion-based multilevel method to compute multiple solutions~\cite{hao2024companion}, as detailed in \S \ref{cbm}. Additionally, the homotopy grid expansion method is discussed in \S \ref{fin}. 

\begin{figure}[ht]
    \centering
    \includegraphics[width=.8\textwidth]{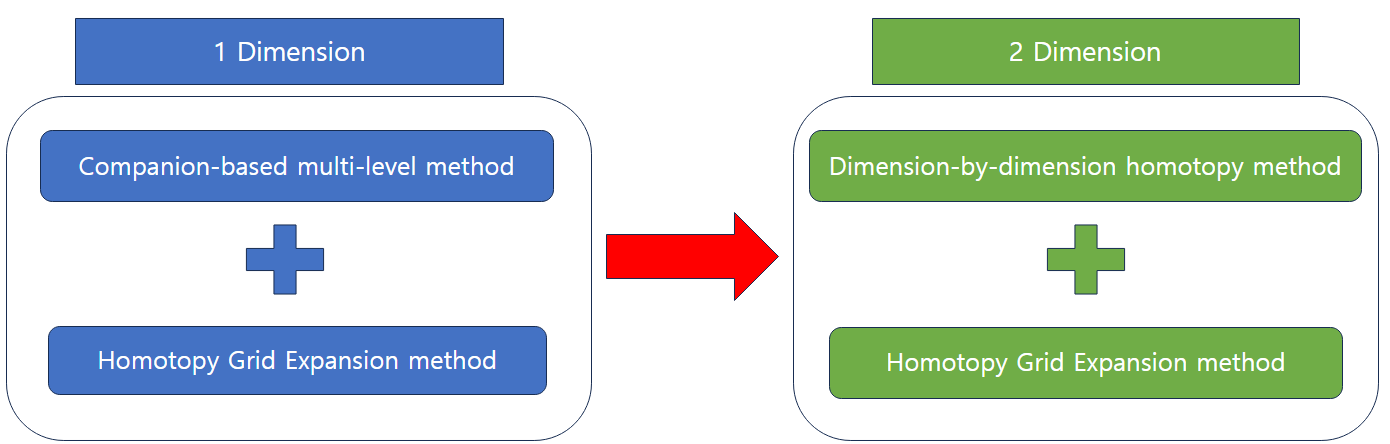}
    \caption{Overview of the numerical methods employed for both 1D and 2D systems. For 1D systems, we first use the companion-based multi-level method and the homotopy grid expansion method to compute multiple solutions. For 2D systems, the solutions obtained from 1D serve as the initial guess, and we then apply the dimension-by-dimension homotopy method and the homotopy grid expansion method to compute the corresponding 2D solutions.}
    \label{illustrate}
\end{figure}

For the two-dimensional case, we develop a dimension-by-dimension homotopy method, which incorporates the homotopy approach on the Laplacian operator, as further described in \S \ref{homod}. The relationship between these numerical methods is illustrated schematically in Fig.~\ref{illustrate}.

While the proposed numerical framework can be naturally extended to Neumann boundary conditions, in this work we focus on the Dirichlet case for clarity and consistency. Our approach is built upon the methodology in~\cite{hao2024companion}, where the companion-based methods are shown to work effectively under different boundary conditions. Therefore, the extension to Neumann boundary conditions can be carried out in a similar manner without fundamental modifications.
Also, it should be remarked that since our solutions decay rapidly to $0$ near the domain
boundaries, we do not, a priori, expect such a modification to provide 
novel waveforms.

\subsection{Companion-based Multi-Level Method in 1D}\label{cbm}

The companion-based multi-level method employed here was originally developed in~\cite{hao2024companion} to compute time-independent solutions of various nonlinear elliptic PDEs with polynomial nonlinearities in both 1D and 2D within a finite element (FEM) framework. While our work uses a finite difference (FDM) discretization, we adapt the key ideas from that FEM-based approach to systematically generate high-quality initial guesses for finer grids and to improve convergence of nonlinear solvers in our 1D Bose–Einstein condensate model.

Using the ansatz in Eq.~(\ref{ansatz}), the relation $\psi_{\mathbb{C}} = c\,\psi_{\mathbb{R}}$ holds, where $c = \tan(\theta)\in\mathbb{R}$ for any $\theta \in [0, 2\pi]$~\cite{charalampidis2018computing}. This property arises from the phase invariance of the equation, which allows the phase factor to be separated, reducing the problem to a real-valued one. It should be noted, however, that while phase invariance exists in higher dimensions for the eGPE, this simplification to a purely real solution is specific to the one-dimensional case. In two dimensions, vortices and multi-vortex configurations with nontrivial phase structures prevent such a reduction.

Assuming $f$ is a nonlinear polynomial function of $\psi$, the system (\ref{main}) can be simplified to a single equation:\begin{equation}\label{1deq}
- \big(\psi_{\mathbb{R}}\big)_{xx} + f_\mathbb{R}(x,\psi_{\mathbb{R}}, c\psi_{\mathbb{R}}) = 0, \quad \mbox{ with } x\in(-d, d).
\end{equation}
To solve Eq. (\ref{1deq}), we utilize the multi-level finite difference method, where the grid consists of $n_l+1$ points, denoted by $l$ as the level index. We define $h_l=\frac{2d}{n_l}$ and grid points $x^l_i=-d+ih_l$, with $i=0,\cdots, n_l$. The numerical solution is denoted as $\Psi_{\mathbb{R}}^{i,l}\approx \psi_{\mathbb{R}}(x^l_i)$ on the $l$-th level grid.

To approximate the second-order derivative, we utilize the central difference method; while the method can be 
easily adapted to higher order stencils, we find this
useful for demonstration purposes. The resulting discretized system of nonlinear equations with Dirichlet boundary condition is represented as ${\bm F}^{l} ({\bm \Psi}_{\mathbb{R}}^{l})=0$, where $\bm\Psi_{\mathbb{R}}^{l}=
(   \Psi_{\mathbb{R}}^{0,l},  \Psi_{\mathbb{R}}^{1,l},\dots,  \Psi_{\mathbb{R}}^{n-1,l}, \Psi_{\mathbb{R}}^{n,l}
)^T
\in \mathbb{R}^{n_l+1}$
and 
\begin{equation}\label{app}
    { F}^{l}_i({\bm \Psi}_{\mathbb{R}}^{l})= 
    \begin{dcases}
    \frac{2\Psi_{\mathbb{R}}^{i,l}-\Psi_{\mathbb{R}}^{i-1,l}-\Psi_{\mathbb{R}}^{i+1,l}}{h_l^2}+f_\mathbb{R}(x_i^{l},\Psi_{\mathbb{R}}^{i,l},c\Psi_{\mathbb{R}}^{i,l}) & 1\leq i\leq n_l-1,\\
    \Psi_{\mathbb{R}}^{i,l}  & i=0 \text{ or } n_l.
    \end{dcases}
\end{equation}

Assuming we have the solution on the $l$-th level grid, denoted as $\bm \Psi_{\mathbb{R}}^{l}\in \mathbb{R}^{n_l+1}$, our goal is to compute the solutions $\bm \Psi_{\mathbb{R}}^{l+1}$ on the $l+1$-th level grid based on $\bm \Psi_{\mathbb{R}}^{l}$.

Given that $n_{l+1}=2n_l$, on the grid points of the $l$-th level, we approximate the solution $\Psi_{\mathbb{R}}^{i,l+1}$ using the solution from the $l$-th level, as follows:
\begin{equation}\label{approxi}
\Psi_{\mathbb{R}}^{2i,l+1}=\Psi_{\mathbb{R}}^{i,l}, \quad i \in \{0,1, \dots, n_l\}.
\end{equation}
For the newly introduced grid points on the $l+1$-th level grid, we solve the following equation to obtain their values:
\begin{equation}
    { F}^{l+1}_{2i+1}({\Psi}_{\mathbb{R}}^{2i+1,l+1})=\frac{2\Psi_{\mathbb{R}}^{2i+1,l+1}-\Psi_{\mathbb{R}}^{2i,l+1}-\Psi_{\mathbb{R}}^{2i+2,l+1}}{h_{l+1}^2}+f_\mathbb{R}(x_{2i+1}^{l+1},\Psi_{\mathbb{R}}^{2i+1,l+1})
\end{equation}
Given that we already approximate $\Psi_{\mathbb{R}}^{2i,l+1}$ and $\Psi_{\mathbb{R}}^{2i+2,l+1}$ using the $l$-th solution using Eq. \eqref{approxi}, ${ F}^{l+1}_{2i+1}({\Psi}_{\mathbb{R}}^{2i+1,l+1})$ becomes a single polynomial equation with a single variable ${\Psi}_{\mathbb{R}}^{2i+1,l+1}$. Specifically, we can express it as:
\begin{equation}\label{poly}
    { F}^{l+1}_{2i+1}({\Psi}_{\mathbb{R}}^{2i+1,l+1}):=\sum_{k=0}^{m} a^{{i,k}}_{l+1} (\Psi_{\mathbb{R}}^{2i+1,l+1})^k=0
\end{equation}
where $a^{{i,k}}_{l+1}$ denotes the coefficient of the $k$-th degree monomial. To compute all the solutions of (\ref{poly}), we compute all the eigenvalues of the following companion matrix:
\begin{equation}\label{companion}
 \left[ \begin{array}{ccccc}
0 & 0 &  \ldots & 0& -a^{{i,0}}_{l+1}/a^{{i,m}}_{l+1}  \\
1 & 0 & \ldots & 0& -a^{{i,1}}_{l+1}/a^{{i,m}}_{l+1}  \\
0 & 1 &  \ldots & 0& -a^{{i,2}}_{l+1}/a^{{i,m}}_{l+1}  \\ 
\vdots & \vdots &\ddots & \vdots  &  \vdots \\ 
0 & 0 & \ldots & 1& -a^{{i,m-1}}_{l+1}/a^{{i,m}}_{l+1}
\end{array} \right].
\end{equation}

By solving polynomials using the companion matrix on the newly added grid points at level $l+1$, we obtain high-quality initial guesses for the solutions at this level. Combining all these solutions, we then apply Newton's method, using these combinations as initial guesses to compute the solutions $\bm{\Psi}_{\mathbb{R}}^{\,l+1}$. A detailed algorithm is provided in Algorithm~\ref{alg:cbmfem}. 

It should be noted, however, that the efficiency of the proposed algorithm can degrade as the number of initial guesses grows exponentially with increasing $l$. To mitigate this, we employ filtering conditions to discard unreasonable or redundant solutions:

\begin{itemize}
    \item \textbf{Locality condition:} The initial guess is assumed to be close to $\bm{\Psi}_{\mathbb{R}}^{l}$ in terms of the residual, i.e.,
    \begin{equation}
        \| \bm F^{\,l+1}(\bm \Psi_{\mathbb{R}}^{\,l+1})\|_{2} < C_1 \, \| \bm F^{\,l}(\bm \Psi_{\mathbb{R}}^{\,l})\|_{2};
    \end{equation}

    \item \textbf{Convergence condition:} We impose a convergence estimate on the initial guess:
    \begin{equation}
        \| \bm F^{\,l+1}(\bm \Psi_{\mathbb{R}}^{\,l+1})\|_{2} < C_2 h^2;
    \end{equation}

    \item \textbf{Boundedness condition:} The initial guess is assumed to be bounded:
    \begin{equation}
        \| \bm \Psi_{\mathbb{R}}^{\,l+1} \|_{\infty} < C_3.
    \end{equation}
\end{itemize}

Here, $\|\cdot\|_2$ denotes the $L^2$ norm and $\|\cdot\|_{\infty}$ the $L^{\infty}$ norm. The constants $C_1, C_2$, and $C_3$ are chosen empirically based on numerical experimentation. Specifically, $C_1$ is calibrated to ensure a sufficient decrease in the residual, $C_2$ reflects the expected order of accuracy of the discretization (second order in $h$), and $C_3$ serves as a safeguard against spurious blow-up in the solution norm. These filtering conditions are motivated by the underlying theoretical framework, while the specific constants are determined empirically; for further details on the theory, see~\cite{hao2024companion}.

\begin{algorithm}
\caption{Companion-based multi-level method in 1D}\label{alg:cbmfem}
\textbf{Input:} Numerical solutions $\bm \Psi_{\mathbb{R}}^{l}\in \mathbb{R}^{{n_l+1}}$ on the $l$-th level. 

\textbf{Output:} $\bm \Psi_{\mathbb{R}}^{l+1}\in \mathbb{R}^{{n_{l+1}+1}}$
\begin{algorithmic}[1]
\State Initialize: $\Psi_{\mathbb{R}}^{2i,l+1}= \Psi_{\mathbb{R}}^{i,l},\, i=0,\cdots,n_l$
\While{$i < n_l$}
    \State Construct the companion matrix using polynomial ${ F}^{l+1}_{2i+1}({\Psi}_{\mathbb{R}}^{2i+1,l+1})=\sum_{k=0}^{m} a^{{i,k}}_{l+1} (\Psi_{\mathbb{R}}^{2i+1,l+1})^k$.
    \State Compute all eigenvalues of the companion matrix
\EndWhile
\State Combine all the solutions of $\Psi_{\mathbb{R}}^{2i+1,l+1}$ as initial guesses.
\State Employ Newton's method to compute $\bm \Psi_{\mathbb{R}}^{l+1}$ using the initial guesses by solving $\bm F^{l+1}(\bm \Psi_{\mathbb{R}}^{l+1})=0$.
\end{algorithmic}
\end{algorithm}

\subsection{Homotopy Grid Expansion method}\label{fin}

In this section, we introduce the homotopy grid expansion method under Dirichlet boundary conditions with zero boundary values, which systematically increases the number of grid points~\cite{allgower2012numerical,seydel2009practical}. Given the solutions at the $l$-th level, this method computes solutions on the $(l+1)$-th level.

While the companion-based multi-level method is effective on coarse grids, its applicability to finer grids is limited because the number of possible solution combinations grows exponentially with grid refinement. Once a sufficiently fine grid is reached, we transition to the homotopy grid expansion method, which requires only polynomially many combinations, providing a practical mechanism to obtain solutions on finer grids.

Specifically, we define the following homotopy function \cite{hao2022adaptive,hao2014bootstrapping}:
\begin{equation}\label{hgp}
    \bm{H}(\bm\Psi,s)= s \bm{F}^{l+1}(\bm\Psi)+ (1- s )\bm{\tilde{F}}^{l+1}(\bm\Psi)=0,
\end{equation}
where $\bm\Psi\in \mathbb{R}^{n_{l+1}+1}$ is the unknown variable, $s\in[0,1]$ is the homotopy parameter, $\bm{F}^{l+1}(\bm\Psi)=0$ is the target system which ${\bm \Psi}^{l+1} $ satisfies, and $\bm{\tilde{F}}^{l+1}(\bm\Psi)=0$ denotes the starting system defined as:

\begin{equation}
{F}^{l}_i(\bm \Psi)=
\begin{dcases}
\frac{2\Psi^{i}-\Psi^{i-1}-\Psi^{i+1}}{h_l^2}+f(x^l_i,\Psi^{i})&\mbox{if } 1< i <  n_l,
\\\Psi^{i} &\mbox{if } i =0 \hbox{~or~} n_l.
\end{dcases}
\end{equation}

\begin{equation}
\tilde{F}^{l+1}_i(\bm \Psi)=
\begin{dcases}
\frac{2\Psi^{i}-\Psi^{i-2}-\Psi^{i+2}}{h_l^2}+f(x^l_i,\Psi^{i})&\mbox{if } 1< i <  2n_l-2,
\\\Psi^{i} &\mbox{if } i =0 \hbox{~or~} 2n_l-1,
\\\frac{2\Psi^{i}-\Psi^{i+2}}{h_l^2}+f(x^l_i,\Psi^{i})&\mbox{if } i =1,
\\\frac{2\Psi^{i}-\Psi^{i-2}}{h_l^2}+f(x^l_i,\Psi^{i})&\mbox{if } i = 2n_l-2.
\end{dcases}
\end{equation}

In practice, the two coarse-grid solutions are merged by alternating their entries, and the resulting vector is treated as the initial guess for the finer grid. The auxiliary system $\tilde{F}^{\,l+1}$ is defined so that this alternately combined vector is an exact solution at $s=0$, providing a consistent starting point for homotopy continuation. In this way, known coarse-level solutions are effectively recycled to generate fine-grid initial guesses.

More precisely, the starting system solution is constructed as:
\begin{equation}\label{construct}
{\bm{\Psi}}=\big(
   0,{\Psi}^{1,l}_\mathbb{R}, \tilde{\Psi}^{1,l}_{\mathbb{R}},{{\Psi}}^{2,l}_\mathbb{R},\tilde{\Psi}^{2,l}_{\mathbb{R}}, \cdots,{{\Psi}}^{n_l-1,l}_\mathbb{R}, \tilde{\Psi}^{n_l-1,l}_{\mathbb{R}}, 0
\big)^T,
\end{equation}
where $\tilde{\bm \Psi}^{\,l}_{\mathbb{R}}$ and $\bm{\Psi}^{\,l}_\mathbb{R}$ are two distinct solutions at the $l$-th level, satisfying $\bm F^{\,l}(\tilde{\bm \Psi}^{\,l}_{\mathbb{R}}) = \bm F^{\,l}(\bm{\Psi}^{\,l}_{\mathbb{R}})=0$. By tracking the homotopy system~\eqref{hgp} from $s=0$ to $s=1$, we obtain the solution on the $(l+1)$-th level grid, $\bm{\Psi}^{\,l+1}_\mathbb{R}$, satisfying $\bm{F}^{\,l+1}(\bm{\Psi}^{\,l+1}_\mathbb{R})=0$.

\begin{figure}[ht]
    \centering
    \includegraphics[width=.8\textwidth]{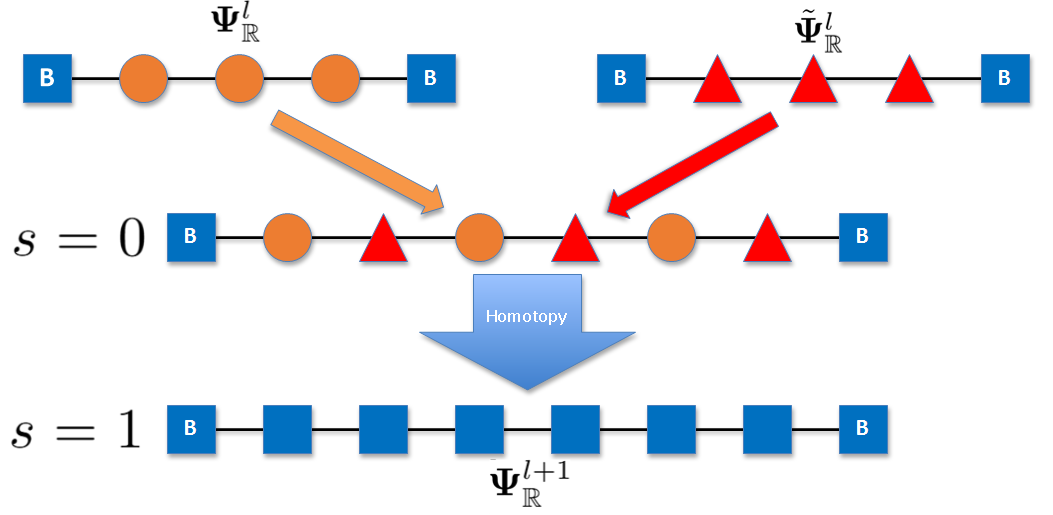}
    \caption{Illustration of the homotopy grid expansion method. Starting from coarse-grid solutions $\tilde{\bm \Psi}^{\,l}_{\mathbb{R}}$ and ${\bm \Psi}^{\,l}_{\mathbb{R}}$, the solutions are combined and the homotopy system is tracked from $s=0$ to $s=1$ to obtain the fine-grid solution ${\bm \Psi}^{\,l+1}_{\mathbb{R}}$. Here, ``B" denotes boundary points.}
    \label{explain}
\end{figure}

An illustration of the homotopy grid expansion method is shown in Fig.~\ref{explain}, and a detailed algorithm is provided in {\bf Algorithm~\ref{alg:main}}.

\begin{algorithm}
\caption{Homotopy Grid Expansion method}\label{alg:main}

{\bf Input:} The solutions  ${{\Psi}}^l_{\mathbb{R}}, \tilde{{\Psi}}^l_{\mathbb{R}}\in  \mathbb{R}^{n_l+1} $ on $l$-th level grid

{\bf Output:} ${{{\Psi}}}^{l+1}_{\mathbb{R}}$ on $l+1$-th level grid
\begin{algorithmic}[1]

\State Initialize $s= 0$; 
\State Initialize $\bm \psi (0)$ based on  ${{\Psi}}^l_{\mathbb{R}}, \tilde{{\Psi}}^l_{\mathbb{R}}$;

\While{$s \le 1$}
 \If{$s==1$}
    \State \textbf{break}

    \ElsIf{$s+ \Delta s \le 1$ }
    \State $s = s+ \Delta s$ 
    \Else
\State $s = 1$ 
    \EndIf

    \State Solve $\bm{H}(\bm\Psi,s)=0$ using  Newton's method
   
\EndWhile
\end{algorithmic}
\end{algorithm}

\subsection{Dimension-by-dimension homotopy method}\label{homod}
We develop a dimension-by-dimension homotopy method to compute 2D solutions based on 1D solutions. Consider a 2D square domain $[-d,d]\times[-d,d]$ with a uniform Cartesian grid defined by $x_i = -d + i h_x$, $y_j = -d + j h_y$ for $i=0,\dots,n_x$ and $j=0,\dots,n_y$, where $h_x = \frac{2d}{n_x}$ and $h_y = \frac{2d}{n_y}$. The numerical solutions are denoted as
\[
\Psi_\mathbb{R}^{i,j} \approx \mathrm{Re}(\psi(x_i,y_j)), \quad 
\Psi_\mathbb{C}^{i,j} \approx \mathrm{Im}(\psi(x_i,y_j)).
\]
For convenience, we represent the solutions as
\[
\bm{\Psi} = \begin{bmatrix} \bm{\Psi}_\mathbb{R}, \bm{\Psi}_\mathbb{C} \end{bmatrix}, \quad
\bm{\Psi}_\mathbb{R} = \begin{bmatrix} \Psi_\mathbb{R}^{i,j} \end{bmatrix}, \quad
\bm{\Psi}_\mathbb{C} = \begin{bmatrix} \Psi_\mathbb{C}^{i,j} \end{bmatrix}.
\]
Next, we define the homotopy function 
$
    \bm F({\bm{\Psi}} , \lambda)=
    \begin{bmatrix}
       F^{i,j}_\mathbb{R}({\bm{\Psi}} , \lambda), 
       F^{i,j}_\mathbb{C}({\bm{\Psi}} , \lambda)
    \end{bmatrix}
$ and 
\begin{equation}
F^{i,j}_\mathbb{R}({\bm{\Psi}}, \lambda)= 
    \begin{dcases}
        {\Psi}^{i,j}_\mathbb{R} \quad \quad  \mbox{if } i=0,n_x \text{ or } j =0,n_y,\\
 \frac{2{\Psi}^{i,j}_\mathbb{R}-{\Psi}^{i-1,j}_\mathbb{R}-{\Psi}^{i+1,j}_\mathbb{R}}{h_x^2}+ \lambda\Big(   
    \frac{2{\Psi}^{i,j}_\mathbb{R}-{\Psi}^{i,j-1}_\mathbb{R}-{\Psi}^{i,j+1}_\mathbb{R}}{h_y^2}\Big)+
f_\mathbb{R}(\sqrt{(x^i)^2+\lambda (y^j)^2},{\Psi}^{i,j}_\mathbb{R},{\Psi}^{i,j}_\mathbb{C}) &\mbox{otherwise } ,
    \end{dcases}
\end{equation}
\begin{equation}
F^{i,j}_\mathbb{C}({\bm{\Psi}}, \lambda )= 
    \begin{dcases}
        {\Psi}^{i,j}_\mathbb{C} \quad \quad  \mbox{if } i=0,n_x \text{ or } j =0,n_y,\\
 \frac{2{\Psi}^{i,j}_\mathbb{C}-{\Psi}^{i-1,j}_\mathbb{C}-{\Psi}^{i+1,j}_\mathbb{C}}{h_x^2}+ \lambda \Big(   
    \frac{2{\Psi}^{i,j}_\mathbb{C}-{\Psi}^{i,j-1}_\mathbb{C}-{\Psi}^{i,j+1}_\mathbb{C}}{h_y^2}\Big)+
f_\mathbb{C}(\sqrt{(x^i)^2+\lambda (y^j)^2},{\Psi}^{i,j}_\mathbb{R},{\Psi}^{i,j}_\mathbb{C}) &\mbox{otherwise } .
    \end{dcases}
\end{equation}
When $\lambda = 1$, $\bm{F}(\bm{\Psi},1)$ corresponds to the full 2D system, which is the target system we aim to solve. When $\lambda = 0$, the system reduces to decoupled 1D problems along the $x$-direction, which serve as the starting system. In this case, we construct initial guesses using the 1D solutions: for each $y_j$, the real part $\psi_\mathbb{R}$ is given by the 1D solution, and the imaginary part is set as $\psi_\mathbb{C} = \tan(\theta_k) \psi_\mathbb{R}$ with $\theta_k = \frac{2\pi}{n_\theta} k$. Consequently, the total number of start-system solutions is $S^{n_y} n_\theta$, where $S$ denotes the number of 1D solutions.
By tracking $\lambda$ continuously from $0$ to $1$ \cite{hao2022adaptive,hao2020adaptive}, the 2D solutions are obtained by gradually deforming the decoupled 1D solutions into fully coupled 2D solutions.

In cases where $n_y$ is large, the computational cost grows exponentially. To mitigate this, we start with a small $n_y$ and employ the homotopy grid expansion method to transition to finer grids. Moreover, instead of considering all possible phase angles, we restrict attention to a small subset of representative values (e.g., $\theta = 0, \pi/2, \pi$), which capture the relevant symmetry classes, such as vortical and stripe states.

We conclude this section by highlighting a key advantage of the proposed approach. Our method provides multiple strategies to generate good initial guesses, which together allow us to efficiently obtain a large number of solutions. Moreover, the approach allows multiple solutions to be identified in parallel in a single computation. This combination of robust initial guess generation and parallel computation makes the method particularly effective in settings where a wide range of known solutions is not available, offering a broadly applicable framework for exploring nonlinear PDEs beyond the specific model considered here.

\section{Numerical Results}\label{sec:Results}

\subsection{Existence and Stability of Waveforms in 2D Confined Settings with Competing Nonlinearities}

The system that has recently attracted significant interest, both theoretically and experimentally~\cite{luo2021new}, involves an interplay between attractive and repulsive effects. In 1D, repulsion is provided by the mean-field cubic nonlinearity, while attraction arises from the quadratic Lee-Huang-Yang correction to the mean-field~\cite{petrov2016ultradilute,tylutki2020collective}. In 2D, a logarithmic nonlinearity effectively captures both effects~\cite{luo2021new}, yielding the governing equation
\begin{equation}\label{ex2}
i \partial_t \Psi = -\frac{\triangle \Psi}{2} + |\Psi|^2 \Psi \ln(|\Psi|^2) + V(r) \Psi, \quad \mbox{in } D \times \mathbb{R}^+,
\end{equation}
where $V(r) = \frac{\Omega^2 r^2}{2}$ is the normalized parabolic trap, as typically used in BEC experiments~\cite{kevrekidis2015defocusing,PitaevskiiStringari2016}.  

Using the standard stationary ansatz from Eq.~\eqref{ansatz}~\cite{alfimov2007nonlinear}, we obtain the time-independent equation
\begin{equation}
0 = -\frac{\triangle \psi}{2} + |\psi|^2 \psi \ln(|\psi|^2) + \frac{\Omega^2 r^2}{2} \psi - \mu \psi,
\end{equation}
with homogeneous Dirichlet boundary conditions.

Our main goal is to explore a diverse set of solutions, including non-topological and vortical waveforms (such as quantum droplets), using the companion-based multi-level method in 1D, the homotopy grid expansion, and the dimension-by-dimension homotopy method, as illustrated in Fig.~\ref{illustrate}.  

To employ the companion-based multi-level method in 1D, it is necessary to reduce the 2D problem to a 1D setting via a dimension restriction:
\begin{equation}\label{1d_re}
i \partial_t \Psi = -\frac{\Psi''}{2} + |\Psi|^2 \Psi \ln(|\Psi|^2) + V(x) \Psi, \quad \text{in } (-d,d) \times \mathbb{R}^+.
\end{equation}
Here $'$ denotes the derivative with respect to a single (in 
this case the $x$) variable.
This dimension restriction differs from the standard 1D quadratic-cubic model considered in prior studies~\cite{katsimiga2023solitary,luo2021new,saqlain2023dragging,tylutki2020collective}. Here, our focus is on 2D waveforms, and thus we exclusively use the restricted equation.  

It is important to note that the solutions of Eq.~\eqref{1d_re} are {\it by construction} transversely homogeneous (i.e., $y$-independent) solutions of the full 2D model. Correspondingly, the 1D potential $V(x)$ represents the 2D problem under a 1D parabolic trap. The stability of these transversely homogeneous solutions is an interesting question on its own. Recent works have shown that generic transverse instabilities are expected~\cite{Bougas_stability_2024}, though notable exceptions exist, such as kink-like waveforms~\cite{mistakidis2024generictransversestabilitykink}, which may exhibit enhanced transverse stability.

\subsubsection{1D restriction results}\label{sec:1D_R}
\begin{figure}
    \centering
    \includegraphics[width=.9\textwidth]{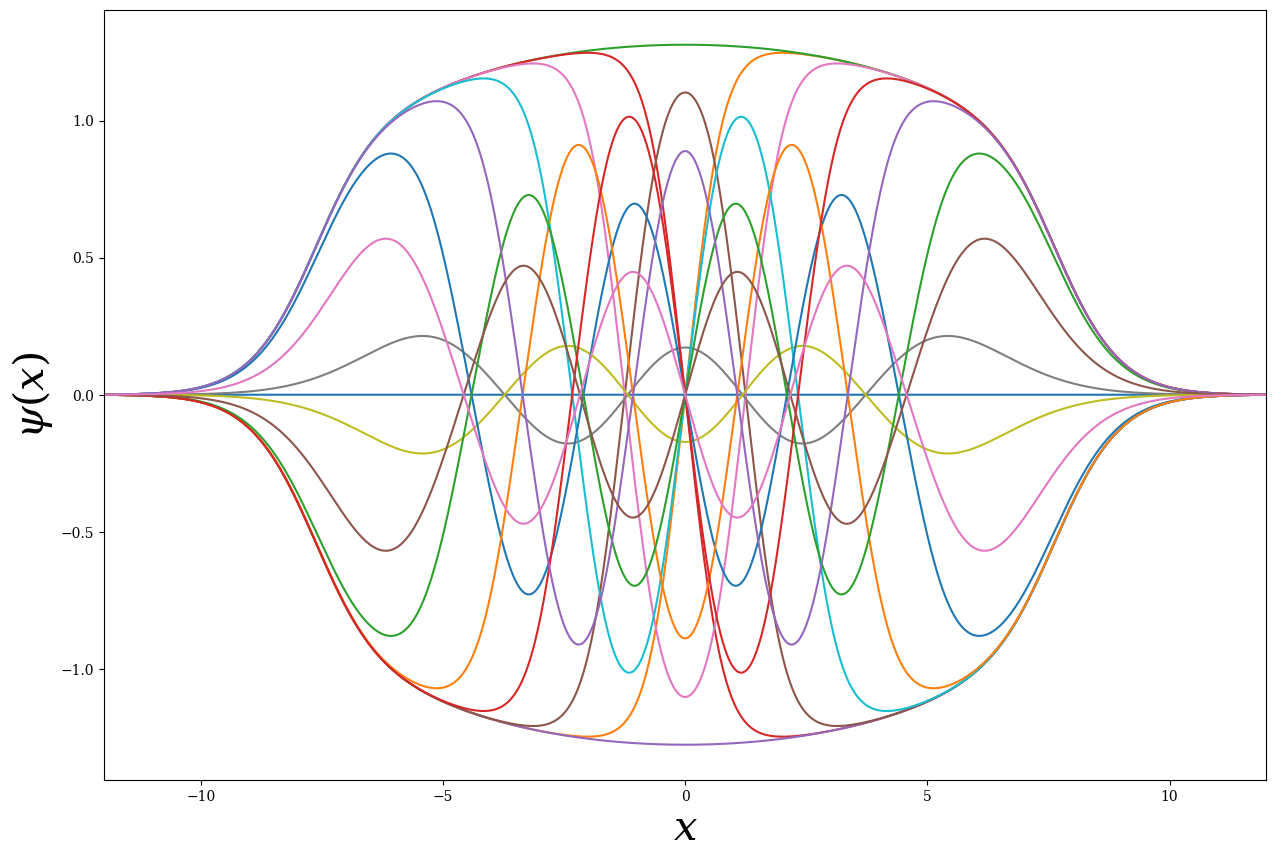}
    \caption{A total of 17 numerical solutions of Eq.~\eqref{1d_re} computed on 1025 grid points over the domain $D = [-12,12]$ with parameters $\mu = 0.8$ and $\Omega = 0.2$, under homogeneous Dirichlet boundary conditions.}\label{1d_2}
\end{figure}

\begin{figure}
    \centering
    \includegraphics[width=.9\textwidth]{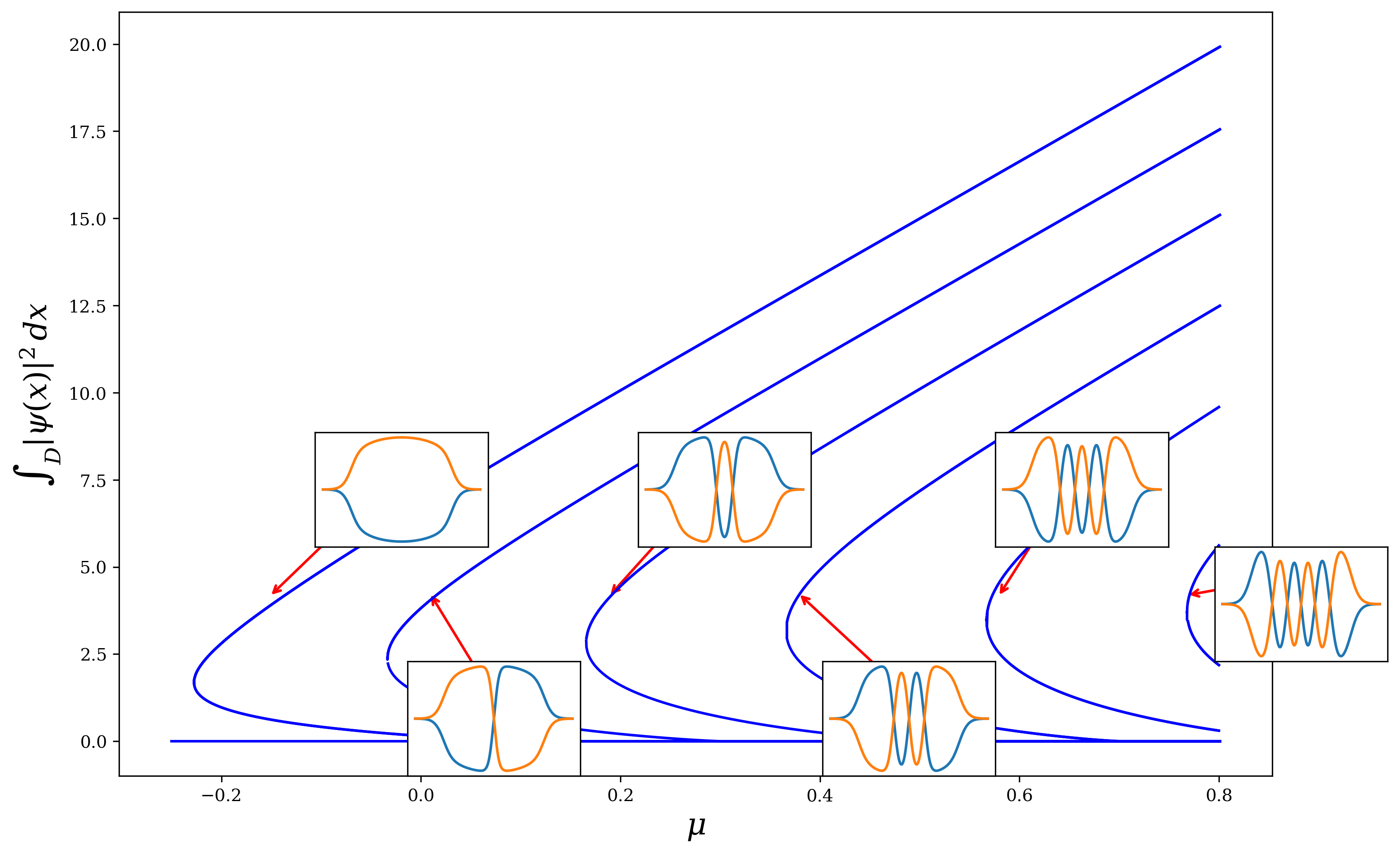}
    \caption{Based on solutions in Fig. \ref{1d_2}, we employed arclength continuation to trace the different solution branches, including those that converge to the trivial state for $\mu = 0.8$. Notably, the first branch encountered corresponds to the ground state, while the second one yields the first excited state, the third  gives the second excited state, and so on.}\label{1dbif_2}
\end{figure}

Given that $\ln(|\Psi|^2)$ entails a non-polynomial nonlinearity, we approximate this term using a polynomial expansion up to the quadratic term. Specifically, we express $\ln(|\Psi|^2)$ as:
\[\ln(|\Psi|^2)= \ln(a)+\sum_{n=1}^{\infty} \frac{(-1)^{n-1}}{n a^n}(|\Psi|^2-a)^n\approx \ln(a)+\sum_{n=1}^{2} \frac{(-1)^{n-1}}{n a^n}(|\Psi|^2-a)^n.\]
Subsequently, we employ the companion-based multi-level method with $a=2^{-k}$, where $k=-1,0,1,\cdots$, until no new solutions can be obtained. Table \ref{1dt} indicates that no new solutions emerge beyond $a=1/8$. We identify 17 solutions on a grid with $N_x=1025$ grid points, as illustrated in Fig. \ref{1d_2}. Subsequently, we utilize the homotopy continuation method to construct bifurcation diagrams of $N=\int |\psi|^2 dx$ versus $\mu$, shown in Fig. \ref{1dbif_2}.

{{Since the companion-based method is inherently designed for polynomial systems, we first approximate the logarithmic nonlinearity with a polynomial to generate initial guesses. The subsequent Newton iterations, however, are performed on the original equations (Eq. (\ref{1d_re})) containing the exact logarithmic term. Thus, the final computed solutions satisfy the original model exactly.}}

In addition to be a mathematically relevant 
bifurcation diagram, it is also a physically meaningful
one, as it represents the (rescaled) atom number as a function
of the model's chemical potential (also, the latter
represents the solution's 
oscillation frequency). It is worthwhile to note that similarly to analogous
1D versions of such diagrams in the cubic defocusing case, e.g., in the works
of~\cite{alfimov2007nonlinear,KIVSHAR2001225}, the ground state bears no dark solitonic excitation (i.e., it
represents a unimodal structure with no zero crossing),
the first excited one features a single dark solitary wave, the second excited
state, two solitary waves, etc.

The resulting bifurcation diagram is very strongly reminiscent of the one presented in the context of the 1D quadratic-cubic nonlinearity, e.g., in~\cite{katsimiga2023solitary} (see Fig.~1 therein). I.e., it involves a single-humped ground state, bifurcating from the ground state of the quantum harmonic oscillator (QHO), a first excited state, involving a single density dip, bifurcating from the first excited state of the QHO, and so on.
{Nevertheless, it is crucial to distinguish our states from those discussed in~\cite{katsimiga2023solitary}. The results of~\cite{katsimiga2023solitary} pertain to one-dimensional solutions of a model with a quadratic-cubic nonlinearity. In contrast, the solutions of Eq.~(\ref{1d_re}) considered here are inherently higher-dimensional, yet
transversely homogeneous structures, whose profiles are obtained by restricting the full higher-dimensional solution to one dimension. Consequently, while the bifurcation diagrams share a similar qualitative structure, the physical origin and mathematical nature of the underlying solutions are distinct.}

\begin{table}[ht!]
    \centering
\begin{tabular}{|l*{4}{|c}r}
\hline$a$ & $\#$ of solutions  & $\#$ of new solutions \\
\hline
$2$ & 5 & \\
\hline
$1$ & 13 & 8\\
\hline
$1/2$ & 15& 2\\
\hline
$1/4$ & 13 &2\\
\hline
$1/8$ & 9 &0\\
\hline
\end{tabular}
\caption{The number of numerical solutions of Eq. (\ref{1d_re}) in 1D by using companion-based multi-level method with the polynomial approximation at $a$
discussed in Section~\ref{sec:1D_R}.} \label{1dt}
\end{table}

\subsubsection{Fully 2D Computations}

Our study is conducted on the domain \(D = (-12, 12)^2\), with parameters \(\Omega = 0.2\) and \(\mu = 0.8\). {We have additionally examined larger computational domains and confirmed that the quantitative features of the solutions remain unchanged, thereby ensuring the robustness of the results presented here.}
From Section~\ref{sec:1D_R}, we obtain 1D solutions using the companion-based multi-level method and the homotopy grid expansion method. Building on these results, we employ the dimension-by-dimension homotopy method to generate effective initial guesses for the 2D case. However, unlike the 1D setting, additional techniques are required to unveil the full structure of 2D solutions. To this end, we apply the arclength continuation method (for a recent account thereof,
see, e.g.,~\cite{DAHLKE2024205}) and also use perturbation techniques ---along suitable eigendirections--- to identify new branches \cite{allgower2012numerical}.

Moreover, we observed that when continuing solutions via arclength, the trajectory does not always approach the trivial state with \(N = \int |\psi|^2 \, dx dy = 0\), but instead may connect to different branches. To clarify the behavior near \(N=0\), we therefore compute the solutions in the linear limit as a reference. As the quantity \(N = \int |\psi|^2 \, dx dy\) becomes small, we approximate solutions using linear eigenfunctions. These linear eigenfunctions can be expressed in terms of Hermite polynomials \(H_m\) and \(H_n\)
in Cartesian form as

\begin{equation}
|m, n \rangle_{c} := H_m(\sqrt{\Omega }x)H_n(\sqrt{\Omega} y)e^{-\Omega r^2 /2},
\end{equation}

where \(m, n \ge 0\) \cite{charalampidis2018computing}. The corresponding eigenvalues are \(\mu = (m+n+1)\Omega\). Here, \(m, n\) are related to the quantum numbers of the quantum harmonic oscillator along the standard Cartesian directions. Additionally, the polar coordinate expression for the linear eigenfunctions reads

\begin{equation}
|k, l \rangle_{p} := r^lL^l_k(\Omega r^2)e^{-\Omega r^2/2}e^{i l \theta},
\end{equation}

where \(L^l_k\) are Laguerre polynomials. The associated eigenvalues are \(\mu = (1+|l|+2k)\Omega\). 
Here, $k$ measures the nodal lines along the radial 
direction, while $l$ denotes the azimuthal wavenumber.
Utilizing these eigenfunctions, we aim to classify our solutions based on their linear limits.

An additional element of our considerations, once the
solutions are identified consists of the spectral stability analysis. The latter reveals that the behavior of the perturbation depends on the nature of the eigenfrequencies \(\omega\), obtained from the linearized stability matrix as in \S \ref{stability}. Specifically:
\newline
\begin{itemize}
    \item A \textbf{purely real} \(\omega\) corresponds to a stable solution.
    \item A \textbf{purely imaginary} \(\omega\) indicates exponential instabilities.
    \item A \textbf{general complex} \(\omega\), i.e.,
    an eigenfrequency quartet leads to oscillatory instabilities.
\end{itemize}

In summary, by combining the above techniques, we are able to uncover new families of 2D solutions and carry out their detailed spectral (and nonlinear) analysis. The results of this investigation are presented below. In particular, near the linear limit, standard continuation techniques starting from linear eigenstates can be applied. By generating good initial guesses with our method and combining multiple strategies, we were able to perform these continuation computations efficiently and systematically, highlighting the practical advantages of our approach.

\subsection{2D Nonlinear Waveforms}
Before presenting further numerical results, we note that the solution branches typically exhibit a turning (bending) behavior. This behavior is induced by the logarithmic term in the nonlinearity, which leads to an effective repulsive interaction for $|\Psi|^2 > 1$ and an effective attractive interaction for $|\Psi|^2 < 1$. This transition gives rise to the observed curvature of the branches and can be seen in most of the solution branches presented below.

\subsubsection{\texorpdfstring{Ground State: Linear limit at $\mu=\Omega=0.2$}{Ground State: Linear limit at mu = Ω = 0.2}}

\begin{figure}
    \centering
\includegraphics[width=.6\textwidth]{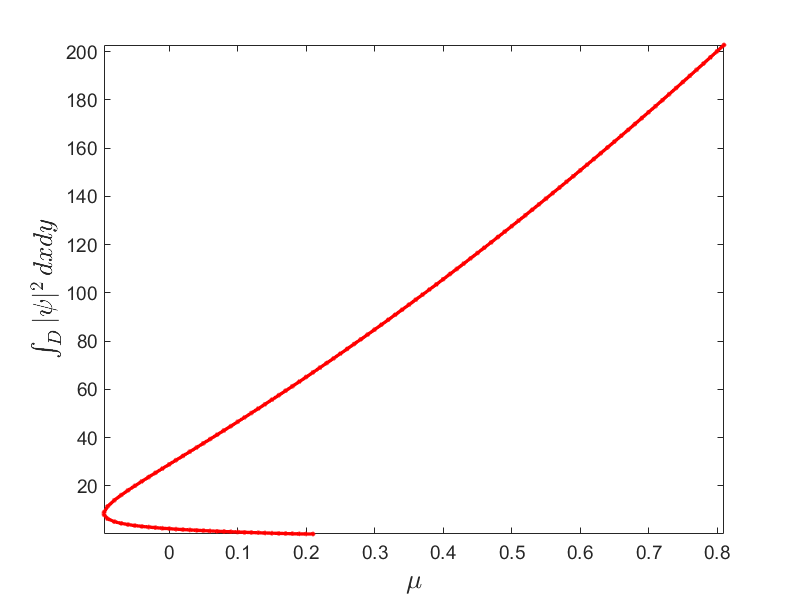}  \includegraphics[width=.4\textwidth]{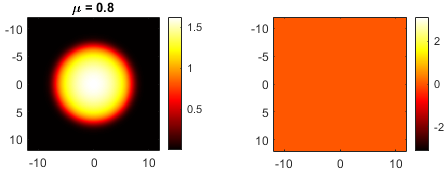}    
    \caption{The first figure visualizes the bifurcation with respect to \( \mu \). The computational domain is chosen as \( D = (-12,12)^2 \) with \( \Omega = 0.2 \). The discretization uses \( N_x = 129 \) and \( N_y = 129 \) grid points. In the lower panels, the left figure shows the density \( |\psi|^2 = \psi_{\mathbb{R}}^2 + \psi_{\mathbb{C}}^2 \), while the right figure displays the corresponding phase angle. The phase angle is defined by \( \theta = \tan^{-1}(\psi_{\mathbb{C}} / \psi_{\mathbb{R}}) \).}\label{mu1}
\end{figure}

Notice that we consider the setting of a parabolic 
trap of frequency $\Omega=0.2$, in line with 
some of the earlier studies, see, e.g.,~\cite{charalampidis2018computing,MIDDELKAMP20111449}.
The simplest state of the system is the ground state, which corresponds to the linear limit state $|0, 0\rangle_{(c)}$ with an eigenvalue of $\mu = \Omega$. This branch is generically stable and does not undergo any bifurcations. As this solution in Fig.~\ref{mu1} is well-known and presents no complex dynamics, it will not be examined further in this work. Importantly, though, we note that similar to 
what is known for the 1D analogue of the problem~\cite{katsimiga2023solitary}
with competing nonlinearities, the relevant
branch in Fig.~\ref{mu1} bends first to the left
and then to the right, after a turning point, yet one that is not associated with
a change in stability.
This reflects the feature of the model that
the dominant nonlinearity is focusing in its
nature for small atom numbers (squared $L^2$ norm
of the solution), while it is defocusing in the 
large atom number, highly nonlinear regime.
Of course, this behavior is fundamentally
different from the monotonic behavior of
the solely cubic (e.g., defocusing) nonlinearity
that was studied in the earlier works mentioned above.

\subsubsection{\texorpdfstring{First Excited State: Linear Limit at $\mu = 2\Omega = 0.4$}{First Excited State: Linear Limit at mu = 2Ω = 0.4}}

For clarity, we include a compact summary table (Table~\ref{Table_mu3}) presenting the main results. The table lists the primary branches, their linear limits, key bifurcation points in $\mu$, and the associated stability transitions, expressed as the number of purely real (i.e., exponentially unstable) eigenvalue pairs. This table is intended as a quick reference; in this subsection, we also discuss oscillatory instabilities and the full eigenvalue spectra in detail.

\begin{figure}
    \centering
\includegraphics[width=.4\textwidth]{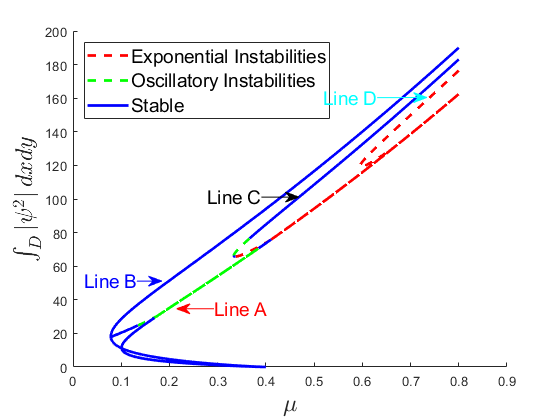}  
\includegraphics[width=.4\textwidth]{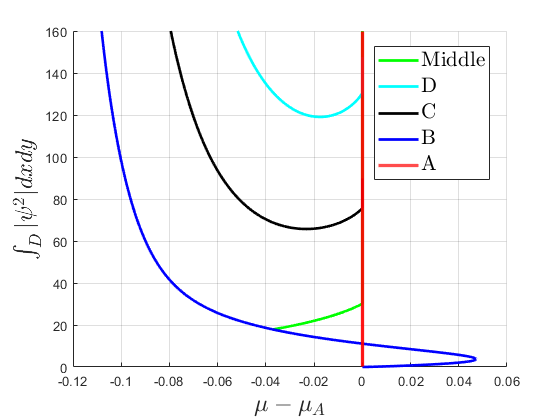}  
\includegraphics[width=1\textwidth]{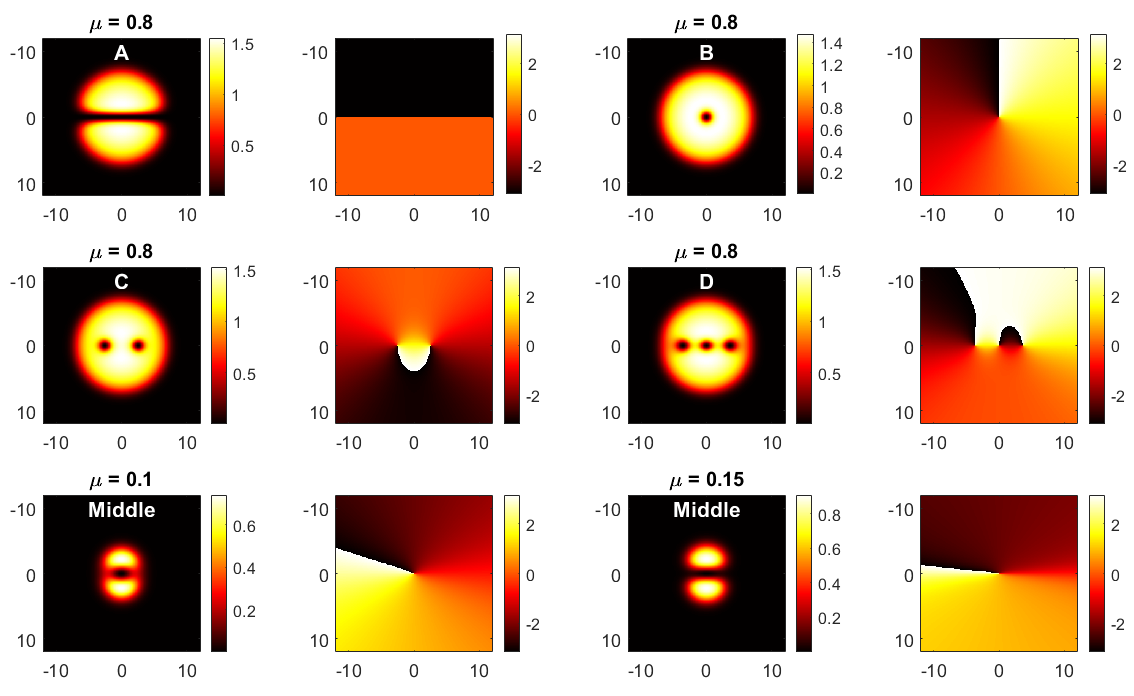}
    \caption{The first figure visualizes the bifurcation with respect to $\mu$. Since the original bifurcation plots are less straightforward to interpret, we produced an alternative visualization in which, for equal values of the $\int_D |\psi|^2dxdy$, the differences in the corresponding chemical potential $\mu$ among the branches and the branch A are plotted on the right for better clarity. We choose $D=(-12,12)^2$ and $\Omega =0.2$. The grid points are $N_x=129$ and $N_y=129$. Here, $|\psi|^2 = \psi_{\mathbb{R}}^2 + \psi_{\mathbb{C}}^2$ is interpreted as the density, and the phase angle is defined by $\theta=\tan^{-1}(\psi_\mathbb{C}/\psi_\mathbb{R})$. A, B, C, and D panels are shown for $\mu=0.8$. The labels of the branches of the top left and top right are designated in terms of their typical profiles in the bottom density and phase (left and right, respectively) plots for each branch. The so-called ``Middle'' branch connects A and B through the green line in the top right panel.
    }\label{mu2}
\end{figure}

The next linear eigenstate occurs when \(m + n = 1\), corresponding to the linear limit \(\mu = 2\Omega\). This scenario is illustrated in Fig.~\ref{mu2}. From a bifurcation perspective, the first significant events emerge at this value of \(\mu\), where two branches bifurcate. Figures~\ref{mu2}~A and~\ref{mu2}~B show the linear limits of the $|1,0\rangle_{(c)}$ and $|0,1\rangle_{(p)}$ states, respectively.  

Physically, the former corresponds to the first excited Cartesian state. For large chemical potentials, the $|1,0\rangle_{(c)}$ branch develops into a dark soliton stripe that is highly susceptible to transverse instabilities~\cite{frantzeskakis2010dark}. In contrast, the $|0,1\rangle_{(p)}$ branch carries a $2\pi$ phase winding, representing a unit-charge vortex. Interestingly, this vortex branch remains generically stable, similar to the cubic defocusing nonlinearity case~\cite{charalampidis2018computing,MIDDELKAMP20111449}, albeit with notable bifurcation differences compared to that classical setting. We now examine these branches and their stability properties in more detail.

\begin{itemize}

    \item \textbf{1-Dark Soliton Stripe ($|1,0\rangle_{(c)}$):} This branch, shown in Fig.~\ref{mu2}~A, is a two-dimensional generalization of the 1D dark soliton state.

    \begin{itemize}
        \item \textbf{Bifurcation Sequence:}
        \begin{itemize}
            \item The first bifurcation associated with this branch occurs around $\mu \approx 0.171$ (Fig.~\ref{mu2}~Middle), after the branch has incurred a (non-stability-changing)
            turning point. This is an unprecedented feature in comparison to the cubic case, as we explain further below. This intermediate state represents a smooth interpolation between the vortex and the dark soliton stripe. As $\mu$ increases, along this
            intermediate branch, the vortex core elongates progressively, gradually transforming into a one-dimensional stripe. In Fig.~\ref{mu2}, the last two panels show how the vortex branch
            ``morphs'' into a dark soliton stripe. Not only is this phenomenology entirely
            absent in the cubic case (where the vortex branch is also always stable
            without incurring any bifurcation), it is also fairly remarkable from a topological
            perspective. This is because a topologically charged configuration (the vortex) 
            gradually deforms
            into a non-topological one (the dark soliton stripe) through the widening and
            eventual disappearance of the phase winding as the vortex core expands
            to reach the condensate edge.
            \item The second bifurcation, at $\mu \approx 0.41$, gives rise to a \textbf{vortex dipole} (Fig.~\ref{mu2}~C). This state consists of a pair of vortices with opposite topological charges ($+1$ and $-1$). The stripe's phase profile undergoes a symmetry-breaking transformation. 
            Here, too, however, this system harbors a significant difference from the cubic case. Namely, the relevant pitchfork bifurcation is {\bf subcritical}. Accordingly, the resulting vortex dipole branch starts out as {\it exponentially unstable}, before it restabilizes upon the relevant turning point, once the defocusing nature of the nonlinearity takes over for higher values of $N= \int |\psi|^2 \, dx dy$.
            \item A third bifurcation takes place at $\mu \approx 0.655$, leading to the formation of a \textbf{vortex tripole} (Fig.~\ref{mu2}~D). This structure comprises three vortices arranged in an alternating charge pattern ($+,-,+$ or $-,+,-$). The vortex tripole represents a higher-order excitation of the stripe, emerging through a subsequent symmetry-breaking bifurcation. These symmetry-breaking events are strongly reminiscent
            of their analogues for the cubic case~\cite{charalampidis2018computing,MIDDELKAMP20111449}
            ---where they have been notably observed in the
            experiments of~\cite{bagnato}---, with the slight
            twist that, per the competing nonlinearity setting herein, the vortex
            dipole and tripole branches ``fork'' backward (i.e., to smaller chemical
            potentials and norms), rather than forward in the immediate vicinity
            of the bifurcation point. 
        \end{itemize}

        \item \textbf{Stability Analysis of the Primary Branch:}
        \begin{itemize}
            \item Starting from the linear limit, the branch is stable. In the vicinity of
            the first bifurcation, the stripe branch develops an oscillatory instability. As the parameter $N= \int |\psi|^2 \, dx dy$ increases, the branch temporarily regains stability just before the second bifurcation. After the second bifurcation, an exponentially unstable eigenmode emerges. The dipole emergence is reminiscent the corresponding bifurcation of the 
            cubic case~\cite{charalampidis2018computing,MIDDELKAMP20111449}, although,
as noted above, the character of the bifurcation changes from supercritical
to subcritical.

            \item \textbf{Remark:} It is worthwhile to recall that the dark soliton
            stripe represents a rotationally invariant configuration. Accordingly, 
            one expects two pairs of eigenvalues lying at the origin of the spectral
            plane, i.e., at $\omega=0$, reflecting the overall phase invariance, as
            well as the rotational invariance of the solution. As a comment aimed
            at numerical practitioners, we should note that
            we have had considerable numerical trouble in preserving the rotational
            invariance eigenpair at the origin. While this 
            (second) eigenfrequency pair is found
            at the origin at low chemical potential, it is progressively seen to depart
            from there (indeed slightly yet progressively so), as the chemical potential
            increases. This is because the solution expands and in a square grid progressively
            ``perceives'' the breaking of the invariance within the numerical grid. 
            We partially amended this by utilizing polar grids. There, the presence of
            a circular domain does not enable an identification of the invariance
            breaking due to perceiving the domain edges. {\it However}, as $N= \int |\psi|^2 \, dx dy$
            increases, the solution populates outermost ``shells'' of the polar
            grid, again ``feeling'' more the lack of perfect rotational invariance
            (due to the higher grid ``coarseness'' at large distances)
            and gradually developing a small nonzero component of the relevant eigenfrequency
            associated with rotation {(see also Appendix \ref{polar_coord}).} We have confirmed by utilizing finer and finer grids
            (for the same domain size) that the relevant eigenfrequency pair approaches
            $\omega=0$, accordingly restoring the symmetry. We have found this aspect
            an extremely subtle one, potentially inhibiting an understanding of the system's
            bifurcations. {Due to this subtle numerical effect affecting the rotational invariance eigenpair, we have omitted 
            any discussion of the eigenvalues near the third bifurcation (Fig.~\ref{mu2}~D), as their apparent 
            computation at large $N$ may be deemed to be less reliable.}         
        \end{itemize}
    \end{itemize}

\item \textbf{Single-Charge Vortex ($|0,1\rangle_{(p)}$):}  
This state, shown in Fig.~\ref{mu2}~B, corresponds to the linear combination $|1,0\rangle_{(c)} + i|0,1\rangle_{(c)}$ in Cartesian coordinates. It represents a robust single-charge vortex that remains stable throughout the considered parameter regime.  

During the bifurcation process of the branch denoted by 
Fig.~\ref{mu2}~Middle, taking place at $\mu \approx 0.078$, and corresponding to the first bifurcation, nearby stable eigenvalues approach the origin, cross it, and then return to the stable side. Importantly, for the vortex branch, this origin crossing occurs entirely within the imaginary axis (i.e., the eigenfrequencies remain real), so the overall stability of the branch is unaffected.  

To clarify the spectral evolution along the intermediate states (i.e., for the branch Fig.~\ref{mu2}~Middle): starting from the vortex branch (Fig.~\ref{mu2}~B), the intermediate ``Middle'' branch has an extra pair of eigenfrequencies at the origin, reflecting its anisotropic nature with ``elongated vorticity'' (see the bottom left panel of Fig.~\ref{mu2}). This branch still enjoys rotational invariance and therefore possesses four eigenfrequencies at $\omega=0$.  

Remarkably, when this branch merges with the dark soliton stripe, a bifurcation occurs, yet no eigenvalue crosses the origin: both the ``Middle'' branch and the stripe branch each retain four eigenfrequencies at the origin, so the stability does not change. No other eigenvalue approaches the origin during this event. While such an event is not forbidden by bifurcation or index theory, it is highly unusual in our experience with similar systems.

    \item \textbf{Main Finding: Continuous Transition.}  
    In summary, a key result of our analysis is the identification of a continuous pathway, shown in Fig.~\ref{mu2}~Middle, that smoothly connects the single-charge vortex branch to the 1-dark soliton stripe branch. This finding is particularly noteworthy, as such a connection was absent in previously studied models with cubic nonlinearities, where distinct branches remained disconnected \cite{charalampidis2018computing,MIDDELKAMP20111449}.
    Hence, while the turning point of the branches
    is natural to expect on the basis of the competing
    nonlinearity, this connection of the vortex and
    the stripe branch is unprecedented to the best
    of our knowledge. This is the ``connecting'' green branch in the top right panel of Fig.~\ref{mu2}.
    In addition to the structural lack of such a precedent, it is
    an especially unusual bifurcation event, given that {\it neither} in the emergence
    of the elongated vorticity Middle branch (from the vortex one), {\it nor} in the collision
    of this Middle branch with the stripe, does a stability change get recorded, per the
    eigenfrequency motions discussed above.

    To summarize the findings of Fig.~\ref{mu2} from a stability perspective, we
    have developed a novel, to our knowledge, form of visualization of the
    relevant bifurcation diagram which is shown in 
    Fig.~\ref{fig:explain}. This is the main workhorse for the presentation of our
    results. The figure shows the bifurcation curves designated by a triplet of
    indices. These three entries indicate, respectively, how many real eigenvalues each branch
    possesses, how many zero eigenvalues and how many imaginary eigenvalues
    {\it close to the origin} (i.e., potentially participating in a bifurcation)
    in the immediate proximity of a bifurcation point. Accordingly, from left
    to right the summary that Fig.~\ref{fig:explain} provides in connection
    to Fig.~\ref{mu2} is that the blue vortex branch used to have 2 imaginary
    eigenvalues before and 2 after the emergence of the rotationally invariant
    (elongated vorticity) Middle green branch. The latter has 4 eigenvalues
    near the origin, as does the stripe branch {\it before and after} their
    collision point in this second very unusual bifurcation event. While Fig.~\ref{mu2} contains the actual computed data, Fig.~\ref{fig:explain} provides a qualitative representation to clarify the connectivity and stability transitions of the different branches.
\end{itemize}

\begin{figure}
    \centering
    \includegraphics[width=0.7\linewidth]{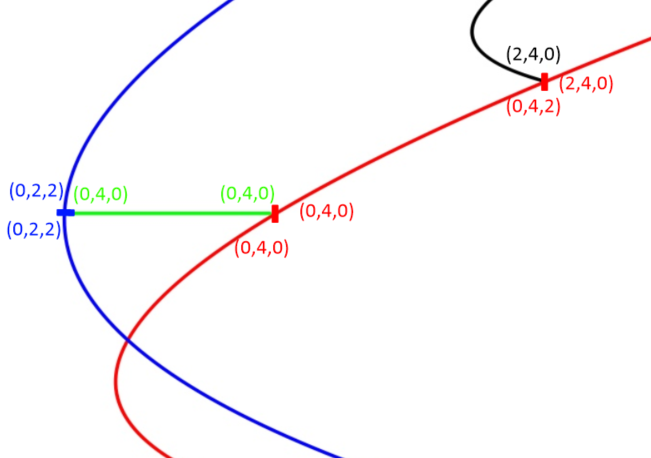}
    \caption{Bifurcation diagram in the $(\mu,\int_D |\psi|^2\,dx\,dy)$ plane associated with Fig.~\ref{mu2}. A simplified version is shown here for clearer visualization. Each tuple $(a,b,c)$ denotes, from left to right, the number of real eigenvalues (i.e., purely imaginary $\omega$ from \S \ref{stability}) $a$, the number of eigenvalues exactly equal to zero $b$, and the number $c$ of eigenvalues that approach the origin from the stable side (imaginary eigenvalues/real eigenfrequencies) as the bifurcation point is approached. At the bifurcation points, these numbers may change, but their total sum $a+b+c$ remains constant before, during and after the bifurcation. For the explanation of the branches and associated stability changes (or lack thereof), see the text.}
    \label{fig:explain}
\end{figure}

\subsubsection{\texorpdfstring{Second Excited State: Linear Limit at $\mu = 3\Omega = 0.6$}{Second Excited State: Linear Limit at mu = 3Ω = 0.6}}

For clarity, we include a compact summary table (Table~\ref{Table_mu3}) that presents the main results for the second excited state as well. {As in the previous case, the table summarizes only the exponential instabilities by listing the number of real unstable eigenvalue pairs associated with each branch. 
Due to the substantially increased complexity of the bifurcation diagram at this excitation level, oscillatory instabilities are not included in this discussion; instead, we distinguish only between exponential instabilities and eigenvalue spectra that remain purely imaginary. 
This choice also allows us to more clearly identify which branch's eigenvalues pass through the origin and which do not, a structural feature that plays an increasingly central role in the organization of the bifurcation diagram.}

{The next linear eigenstate emerges when the condition \(m + n = 2\) is satisfied, leading to the linear limit value \(\mu = 3\Omega\).}
This is analogous to the presentation for the previous state, although 
the dramatically increasing complexity of the bifurcation
diagram should be evident in this case.
 
More specifically, the results based on the parent branches from the linear limit are as follows:

\begin{itemize}
    \item \textbf{From Cartesian Basis States and their Bifurcations}:
        \begin{itemize}
            \item \textbf{Dark Soliton Cross ($|1,1\rangle_{(c)}$)}: this is shown in Fig.~\ref{S_E_1,1}~A. 
            \item \textbf{2-Dark Soliton Stripe ($|2,0\rangle_{(c)}$)}: this branch is shown in Fig.~\ref{S_E_20}~A.
            \item \textbf{From $|2,0\rangle_{(c)}+i|1,1\rangle_{(c)}$}: this is shown in Fig.~\ref{sol_new}~A and is a 
            structure that appears to bear multiple topological charges.
        \end{itemize}
   \item \textbf{From Radially Symmetric Solutions and their Bifurcations}:
        \begin{itemize}
            \item \textbf{Ring Dark Soliton ($|1,0\rangle_{(p)}$)}: This state, also represented as $|2,0\rangle_{(c)} + |0,2\rangle_{(c)}$: see Fig.~\ref{S_E_10p}~A.
            \item \textbf{Vortex Quadrupole ($|2,0\rangle_{(c)}+i|0,2\rangle_{(c)}$)}: this is shown in Fig.~\ref{S_E_10p}~C.
            \item \textbf{Charge-two Vortex ($|0,2\rangle_{(p)}$)}: as depicted in Fig.~\ref{S_E_1,1}~E.
        \end{itemize}
\end{itemize}

Here we analyze the solution branches that emerge from the linear limit at the second excited state. We examine the bifurcation sequences and stability profiles for three primary branches.

\begin{itemize}
    \item \textbf{Branch from $|1,1\rangle_{(c)}$ (Dark Soliton Cross)}
    
    The primary branch (Fig.~\ref{S_E_1,1}~A), originating from the $|1,1\rangle_{(c)}$ state, undergoes a series of bifurcations as $N=\int_D |\psi^2| dx dy$ increases.

    \begin{itemize}
        \item \textbf{Bifurcation Sequence:}
        \begin{itemize}
            \item At $\mu \approx 0.312$, the first bifurcation occurs, creating a new branch (Fig.~\ref{S_E_1,1}~B). {As illustrated in Fig.~\ref{S_E_1,1}, the initial configuration (Fig.~\ref{S_E_1,1}~A) corresponds to a cross state, where two orthogonal dark-soliton stripes intersect. Remarkably, as the parameter is varied, the angle between the initially orthogonal stripes gradually decreases, and the configuration transforms into two nearly parallel dark-soliton stripes oriented along the diagonal direction (Fig.~\ref{S_E_1,1}~B).}

            \item A second bifurcation at $\mu \approx 0.38$ creates a new branch (Fig.~\ref{S_E_1,1}~Middle). As its profile appears to be an interpolation between two primary branches Dark Soliton Cross branch (Fig.~\ref{S_E_1,1}~A) and the Charge-two Vortex branch (Fig.~\ref{S_E_1,1}~E), we refer to it as the 'Middle' branch. Indeed, this Middle
            branch connects these two other branches. {This is again a 
            remarkable feature of the present model: not only is this phenomenology entirely absent in the cubic case, it is also fairly remarkable from a topological perspective. This is because a topologically charged configuration (charge-two vortex) gradually deforms into a non-topological one (the dark soliton cross), leading to the eventual disappearance of the phase winding as the charge-two vortex expands to reach the condensate edge.}
            
            \item A subsequent bifurcation with another branch (Fig.~\ref{S_E_1,1}~C) occurs at $\mu \approx 0.407$. This is a vortex quadrupole
            (with two positive and two negative charges) which appears to bifurcate also from the dark soliton cross. {In the cubic case, this branch appears only from the linear limit. In contrast, in our setting we additionally observe a bifurcation emerging from Fig.~\ref{S_E_1,1}~A, and we identify a corresponding linear-limit branch in Fig.~\ref{S_E_10p}~C, although the latter persists only over a very short parameter interval.}

            \item The final observed bifurcation takes place at $\mu \approx 0.619$ (Fig.~\ref{S_E_1,1}~D).
            This breaks one of the stripes into vortical patterns, while leaving the other one intact. A similar bifurcation was present in the cubic nonlinear 
            case~\cite{charalampidis2018computing}.
        \end{itemize}
        \item \textbf{Stability Analysis of the Primary Branch:}
        \begin{itemize}
            \item At the linear limit, the A branch possesses one pair of unstable eigenvalues. At the first bifurcation (leading to the emergence
            of the solution in Fig.~\ref{S_E_1,1}~B), the branch becomes stable, as the unstable eigenvalue pair disappears. 
            In this subcritical pitchfork bifurcation, the unstable eigenvalue pair
            is inherited by the branch  in Fig.~\ref{S_E_1,1}~B.
            At the second bifurcation (Fig.~\ref{S_E_1,1}~Middle), similarly to the behavior observed for the Middle branch in Fig.~\ref{mu2}, no eigenvalue crossings or changes in stability occur. 
            Indeed, in this case both the parent branch and the daughter branch  bear
            the rotational and phase invariance (hence 4 eigenvalues at the origin) 
            and no other eigenvalue pairs approach the origin near this bifurcation.
            After the third bifurcation (Fig.~\ref{S_E_1,1}~C), an additional unstable eigenvalue pair is introduced. 
            Interestingly, once again (and contrarily to the supercritical features
            of the cubic case of~\cite{charalampidis2018computing,MIDDELKAMP20111449}), the bifurcation
            here is subcritical and leads to an unstable daughter branch C at the fork.
            Following the final bifurcation observed in our computations, 
            i.e., the one leading to the emergence of the branch in Fig.~\ref{S_E_1,1}~D, 
            in addition to its existing instability (through a real pair), the parent branch of Fig.~\ref{S_E_1,1}~A has two more imaginary pairs approaching the origin near this bifurcation
            point.
            As can be inferred from Fig.~\ref{fig:pla}, one of these pairs
            becomes real (while the other stays imaginary) 
            to yield 2 real pairs for branch Fig.~\ref{S_E_1,1}~D, while
            both become real for Fig.~\ref{S_E_1,1}~A in yet another very complex bifurcation
            (that bears both subcritical and supercritical characteristics combined). While Fig.~\ref{S_E_1,1} contains the actual computed data, Fig.~\ref{fig:pla} provides a qualitative representation to clarify the connectivity and stability transitions of the different branches.
        \end{itemize}
    \end{itemize}
\end{itemize}

\begin{figure}[ht!]
\centering\includegraphics[width=0.45\linewidth]{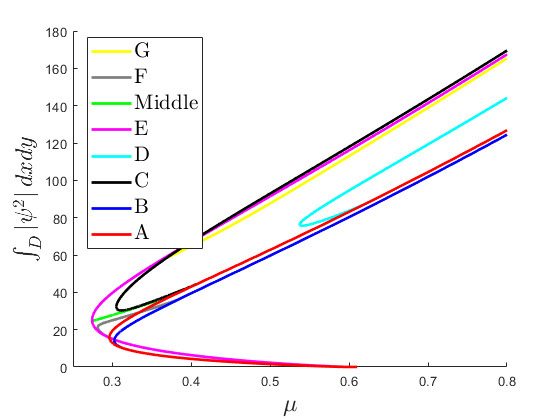}
\includegraphics[width=0.45\linewidth]{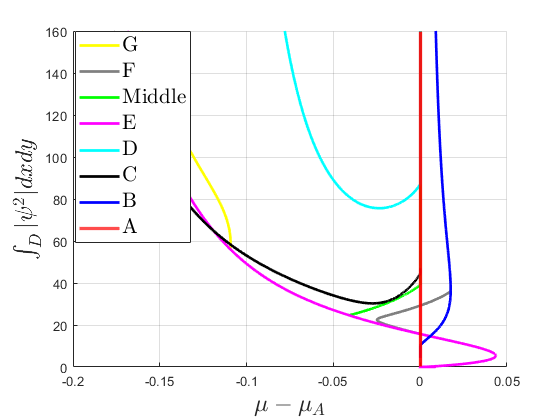}
\includegraphics[width=1\linewidth]{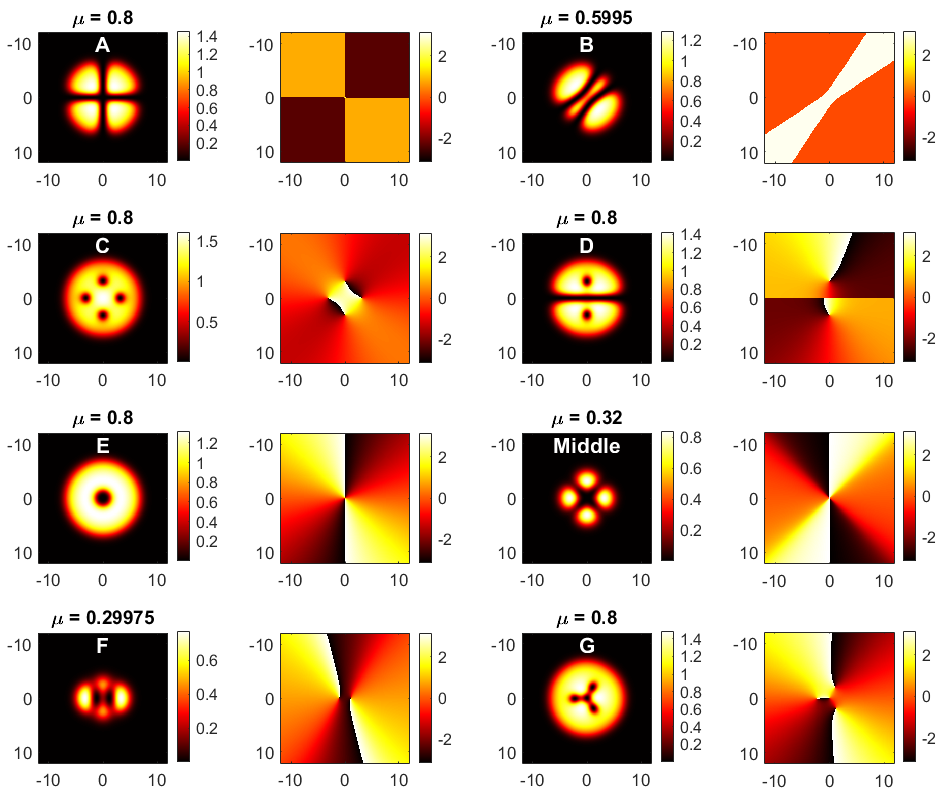}
    \caption{The first figure visualizes the bifurcation with respect to $\mu$ starting with the dark soliton cross $|1,1\rangle_{(c)}$. Since the original plots were difficult to interpret, we produced an alternative visualization in which, for equal values of the $\int_D |\psi|^2dxdy$, the differences in the corresponding $\mu$ among the branches are plotted on the right for better clarity. 
    The visualization and panels are similar to Fig.~\ref{mu2}.}
    \label{S_E_1,1}
\end{figure}
\begin{figure}[ht!]
    \centering
    \includegraphics[width=0.5\linewidth]{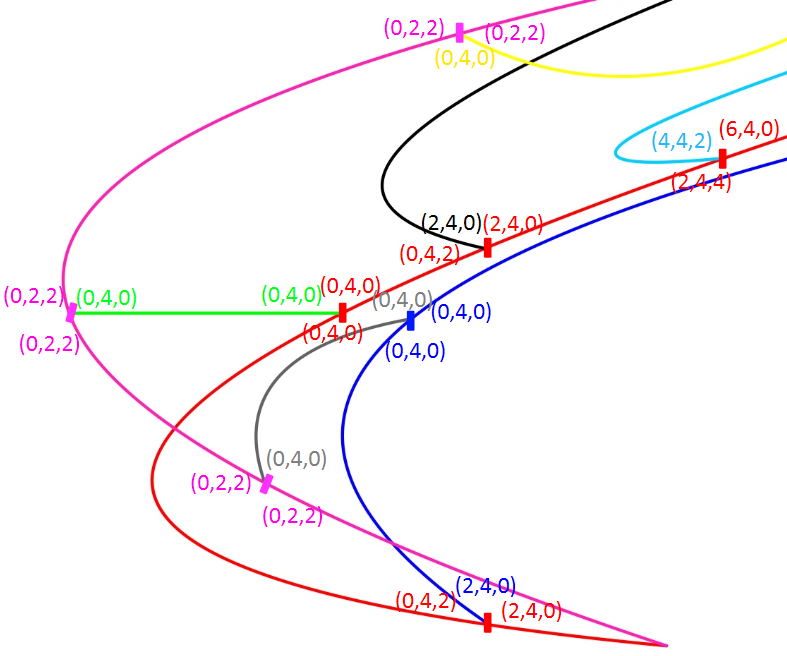}
    \caption{Bifurcation diagram in the $(\mu,\int_D |\psi|^2\,dx\,dy)$ plane with Fig.~\ref{S_E_1,1}. A simplified version is shown here for clearer visualization. Each tuple $(a,b,c)$ denotes, from left to right, the number of unstable eigenvalues (with purely imaginary $\omega$ from \S \ref{stability}) $a$, the number of eigenvalues exactly equal to zero $b$, and the number $c$ of eigenvalues that approach the origin from the stable side as the bifurcation point is approached. During bifurcations, these numbers may change, but their total sum $a+b+c$ remains constant before and after the bifurcation.}
    \label{fig:pla}
\end{figure}

\begin{itemize}
    \item \textbf{Branch from $|0,2\rangle_{(p)}$ (Charge-2 Vortex)} The primary branch (Fig.~\ref{S_E_1,1}~E), originating from the $|0,2\rangle_{(p)}$ state, also undergoes a series of bifurcations as $N=\int_D |\psi^2| dxdy$ increases.
    \begin{itemize}
        \item \textbf{Bifurcation Sequence:}
       \begin{itemize}
            \item At $\mu \approx 0.296$, the first bifurcation occurs, creating a new branch (Fig.~\ref{S_E_1,1}~F) that connects Fig.~\ref{S_E_1,1}~E and Fig.~\ref{S_E_1,1}~B;
            we touched upon the latter in the previous bifurcation sequence.
            \item A second bifurcation at $\mu \approx 0.275$ is associated
            with the branch Fig.~\ref{S_E_1,1}~Middle. As its profile appears to be an interpolation between two primary branches, i.e., the
            dark soliton cross branch of Fig.~\ref{S_E_1,1}~A and the Charge-2 vortex branch of Fig.~\ref{S_E_1,1}~E, we refer to it as the 'Middle' branch.
            \item The final bifurcation that we observed occurs around $\mu \approx 0.358$, giving rise
            to the branch of Fig.~\ref{S_E_1,1}~G, in a way reminiscent of a similar bifurcation 
            in the cubic case, i.e., a charge-2 vortex giving rise to a split state of 3-same charges
            surrounding the origin bearing an opposite charge~\cite{charalampidis2018computing,MIDDELKAMP20111449}.
    \end{itemize}    
        \item \textbf{Stability Analysis of the Primary Branch:}
        \begin{itemize}
            \item From the linear limit, the charge-2 branch exhibits no exponential instability associated
            with a real eigenvalue pair. During the first bifurcation, a single eigenvalue pair 
            approaches and crosses the origin purely along the imaginary axis, similarly to what
            happened with the charge-1 state  in Fig.~\ref{mu2}. Remarkably, here what happens
            is that the charge-2 vortex appears to split into two charge-1 vortices of the branch F.
            An equally remarkable feature pertains to the fact that
            at the second and the final bifurcations observed, a similar eigenvalue crossing near the origin occurs again, whereby the eigenvalue pair crosses through the origin from one to the
            other side of the imaginary axis, without affecting stability. In all 3 cases, the branches
            that arise, the split-charge vortex in F, the hybrid state (between the cross and the charge-2)
            in the Middle and the $3+1$ vortices bear 4 eigenvalues at the origin ---being rotationally
            symmetric---, but not an exponential instability associated with a real eigenvalue pair.
                \item
                {\textbf{Remark:} Regarding Fig.~\ref{S_E_1,1}~B, immediately after the bifurcation in Fig.~\ref{S_E_1,1}~A the branch carries an exponential unstable eigenvalue, whereas in Fig.~\ref{S_E_1,1}~F the spectrum appears qualitatively different. This apparent discrepancy has a clear explanation: along the continuation of this branch, the configuration shown in Fig.~\ref{S_E_20}~B (which corresponds to Fig.~\ref{S_E_1,1}~B) undergoes a further bifurcation, giving rise to the new branch depicted in Fig.~\ref{S_E_20}~C. It is at this additional bifurcation—omitted in the Fig.~\ref{S_E_1,1} sequence—that the eigenvalue structure changes. Because this intermediate bifurcation is not displayed in the Fig.~\ref{S_E_1,1} panels, the spectral difference between Figs.~\ref{S_E_1,1}~B and~F may appear abrupt. For clarity of presentation, we separated these figures, as including the entire bifurcation chain in a single sequence would make the stability changes significantly harder to follow.}
    \end{itemize}
\end{itemize}
\end{itemize}

\begin{itemize}
    \item \textbf{Branch from $|1,0\rangle_{(p)}$ (Ring Dark Soliton)}
    
    The primary branch (Fig.~\ref{S_E_10p}~A), originating from the $|1,0\rangle_{(p)}$ state, undergoes a series of bifurcations as $N=\int_D |\psi^2| dxdy$ increases that we now examine.

    \begin{itemize}
        \item \textbf{Bifurcation Sequence:}
        \begin{itemize}
            \item At $\mu \approx 0.349$, the first bifurcation occurs, giving rise to a new branch (Fig.~\ref{S_E_10p}~B).
            \item A second bifurcation takes place at $\mu \approx 0.337$, where a branch emerging from another primary branch originating at the linear limit (Fig.~\ref{S_E_10p}~C, which is
            reminiscent of a vortex square) merges into Fig.~\ref{S_E_10p}~A .
            \item The next bifurcation is observed around $\mu \approx 0.363$, where the branch (Fig.~\ref{S_E_10p}~B) merges back into the original primary branch (Fig.~\ref{S_E_10p}~A), as shown in Fig.~\ref{S_E_10p}~B. This closed loop scenario is somewhat reminiscent
            of an isola (although it also differs from it).
            \item Around $\mu \approx 0.52$, we find the new branch Fig.~\ref{S_E_10p}~E, which
            is the prototypical bifurcation of vortex hexagon that has been previously
            observed in the cubic case~\cite{charalampidis2018computing,MIDDELKAMP20111449}.
            \item The last bifurcation we observed is around $\mu \approx 0.79$ as shown in Fig.~\ref{S_E_10p}~F. In this case, again reminiscent of the cubic phenomenology, a vortex octagon state
            emerges.
        \end{itemize}
        \item \textbf{Stability Analysis of the Primary Branch:}
        \begin{itemize}
            \item From the linear limit, this ring dark soliton branch exhibits two pairs of exponentially unstable eigenvalues,
          and is unstable  similarly to the cubic case. During the first bifurcation, two pairs of nearby stable eigenvalues cross the origin and become unstable. 
          This leads to 4 real pairs of eigenvalues for the ring dark soliton waveform through this
          supercritical pitchfork.
          At the second bifurcation, a different set of two unstable eigenvalues goes to the stable side
          in another supercritical pitchfork associated with the collision with the
          black branch of Fig.~\ref{S_E_10p}~C. After the third bifurcation, the event closing
          the loop discussed above, the branch becomes fully stable
          spectrally. The last two bifurcations we observed are each associated with the emergence of two pairs of unstable eigenvalues, which subsequently determine the stability changes of the branch.
          It is worthwhile to note that the ring dark soliton branch
          is {\it never} found to be stable in the cubic defocusing
          case of~\cite{charalampidis2018computing,MIDDELKAMP20111449}. While Fig.~\ref{S_E_10p} contains the actual computed data, Fig.~\ref{fig:pla1} provides a qualitative representation to clarify the connectivity and stability transitions of the different branches.
            \item {\textbf{Remark:} However, the behavior of Fig.~\ref{S_E_10p}~D, which connects Fig.~\ref{S_E_10p}~B and C, partially deviates from our expectations. Based on symmetry considerations, one would anticipate two eigenvalue pairs at the origin of the spectral plane (i.e., at $\omega=0$), corresponding to the overall phase invariance and the rotational invariance of the solution. Yet, upon inspecting the eigenvalues in D, we observe six eigenvalues near the origin. It is currently unclear whether these additional eigenvalues are associated with further symmetries of this
            solution or if they simply correspond to small, nonzero eigenvalues. Further investigation is needed to fully interpret this behavior.} Nevertheless, this extra branch D is responsible
            for the restabilization of branch B ---upon their collision---, which, in turn, is
            responsible ---when the loop closes--- for the restabilization of the ring dark soliton branch A.
            We have, nevertheless, once again elucidated the associated highly complex bifurcation diagram.
        \end{itemize}
    \end{itemize}
\end{itemize}

\begin{figure}[ht!]
    \centering    \includegraphics[width=0.45\linewidth]{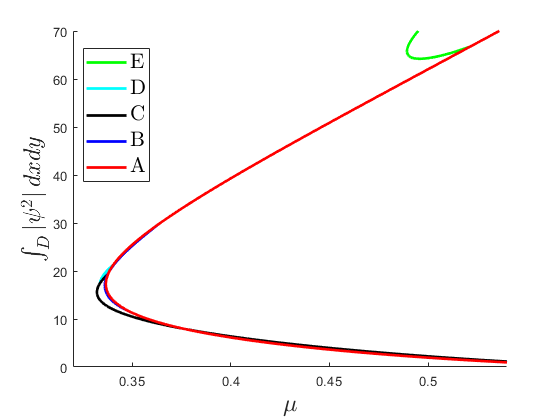}
\includegraphics[width=0.45\linewidth]{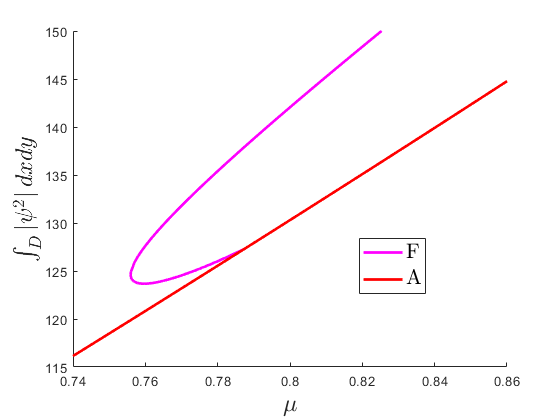}
\includegraphics[width=0.45\linewidth]{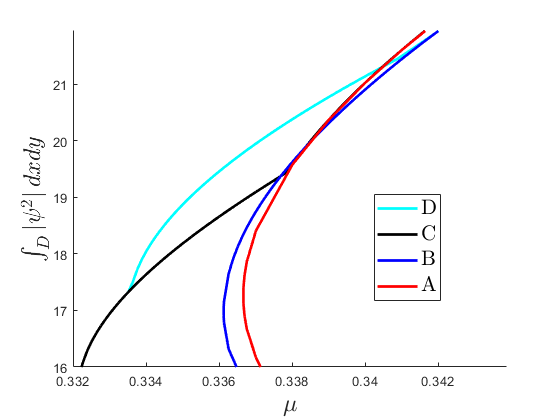}
\includegraphics[width=0.45\linewidth]{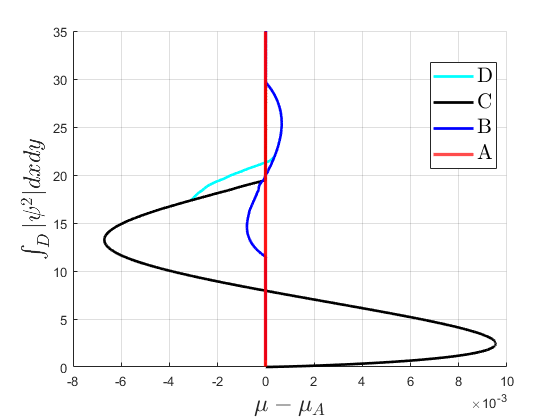}
\includegraphics[width=0.9\linewidth]{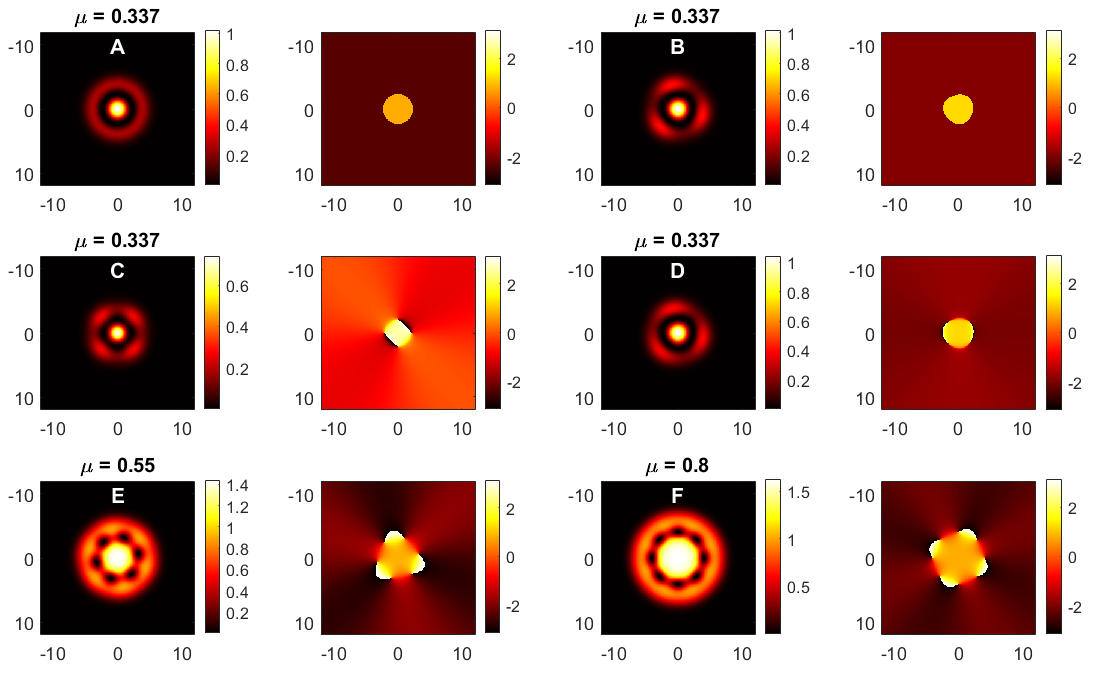}
    \caption{The first figure visualizes the bifurcation with respect to $\mu$, starting with the ring dark soliton $|1,0\rangle_{(p)}$. Since the original plots were difficult to interpret, we produced an alternative visualization in which, for equal values of the $\int_D |\psi|^2dxdy$, the differences in the corresponding $\mu$ among the branches are plotted for better clarity. The visualization and panels are similar to Fig.~\ref{mu2}.}
    \label{S_E_10p}
\end{figure}

\begin{figure}[ht!]
    \centering
    \includegraphics[width=0.5\linewidth]{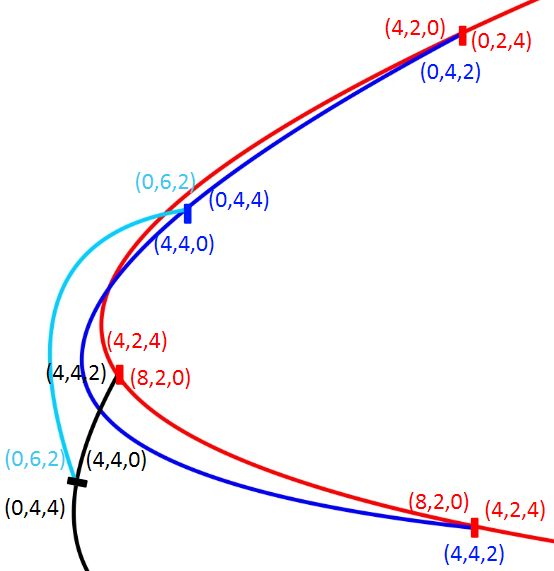}
    \caption{Bifurcation diagram in the $(\mu,\int_D |\psi|^2\,dx\,dy)$ plane with Fig.~\ref{S_E_10p}. A simplified version is shown here for clearer visualization. Each tuple $(a,b,c)$ denotes, from left to right, the number of unstable eigenvalues (with purely imaginary $\omega$ from \S \ref{stability}) $a$, the number of eigenvalues exactly equal to zero $b$, and the number $c$ of eigenvalues that approach the origin from the stable side as the bifurcation point is approached. During bifurcations, these numbers may change, but their total sum $a+b+c$ remains constant before and after the bifurcation.}
    \label{fig:pla1}
\end{figure}
    
\begin{itemize}
    \item \textbf{Branch from $|2,0\rangle_{(c)} + i|0,2\rangle_{(c)}$ (Vortex Quadrupole)}
    
    The primary branch (Fig.~\ref{S_E_10p}~C), originating from the $|2,0\rangle_{(c)} + i|0,2\rangle_{(c)}$ state, merges into the main branch shown in Fig.~\ref{S_E_10p}~A around $\mu \approx 0.337$.
    
    \begin{itemize}
        \item \textbf{Bifurcation Sequence:}
        \begin{itemize}
            \item The first bifurcation occurs around $\mu \approx 0.334$. The corresponding solution (Fig.~\ref{S_E_10p}~D) profile appears to interpolate between the two branches shown in Fig.~\ref{S_E_10p}~C and Fig.~\ref{S_E_10p}~B. Again, here, we have a state that interpolates between a topological one (C) and a non-topological one (B).
            \item The second bifurcation corresponds to a merger with the main branch depicted in Fig.~\ref{S_E_10p}~A at approximately $\mu \approx 0.337$. We already discussed above this
            feature (and its restabilizing effect on the Fig.~\ref{S_E_10p}~A branch).
        \end{itemize}
    
        \item \textbf{Stability Analysis of the Primary Branch:}
        \begin{itemize}
            \item At the linear limit, the branch is stable with no unstable eigenvalues. Following the first bifurcation, it acquires two pairs of unstable eigenvalues in the associated 
            supercritical pitchfork. At the second bifurcation, the branch Fig.~\ref{S_E_10p}~C disappears
            through its collision with the ring dark soliton, leading the ring to acquire anew 2 real
            and 2 imaginary pairs.
        \end{itemize}
    \end{itemize}
\end{itemize}
\begin{figure}[ht!]
    \centering
 \includegraphics[width=0.45\linewidth]{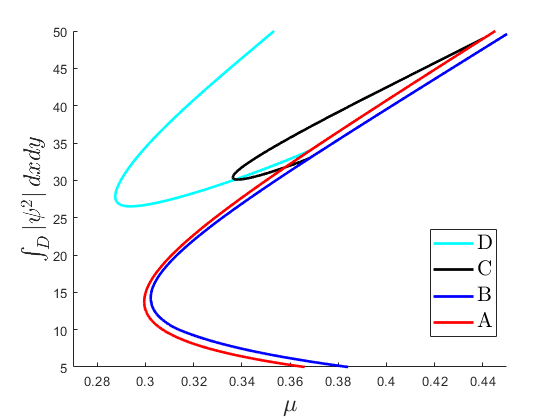}
  \includegraphics[width=0.45\linewidth]{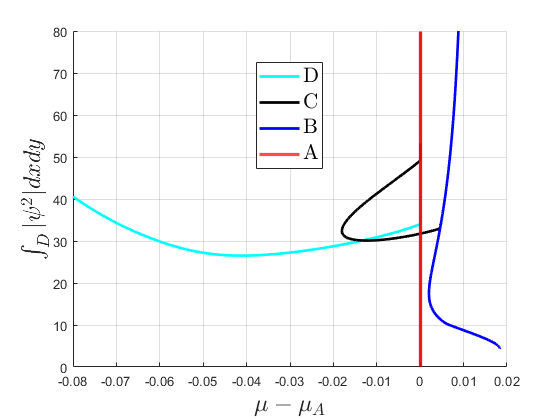}
\includegraphics[width=1\linewidth]{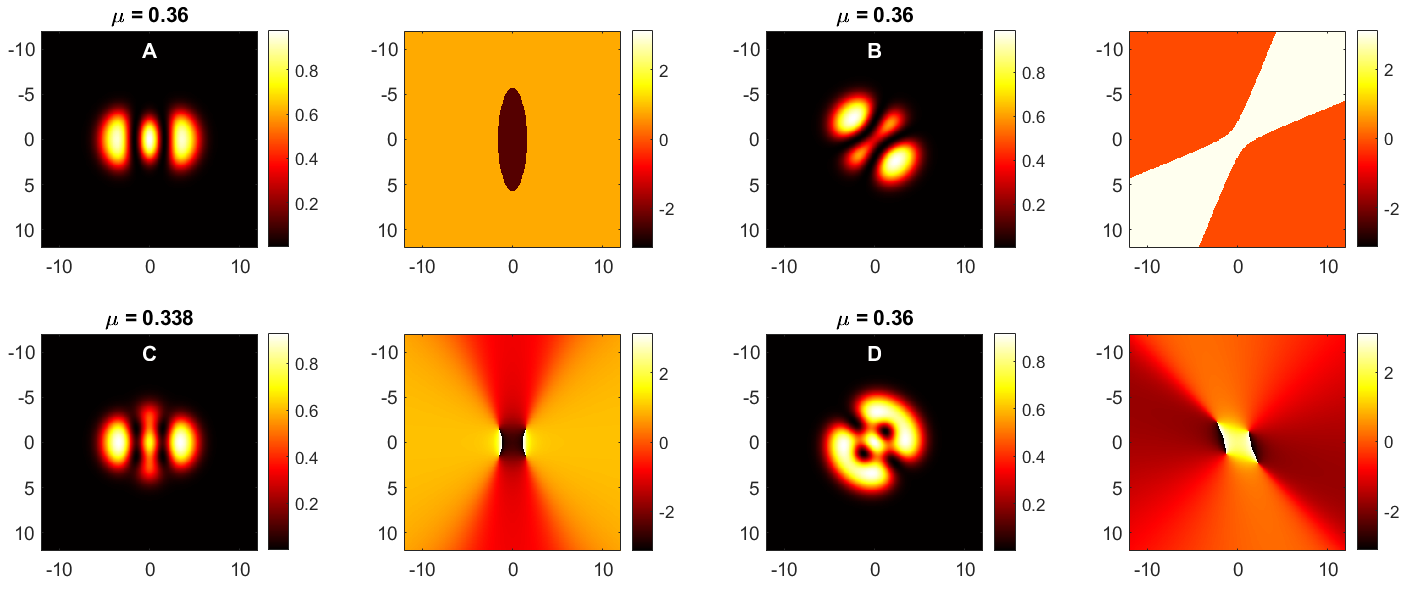}
    \caption{ The first figure visualizes the bifurcations with respect to $\mu$ associated
    with the 2-dark soliton stripe branch $|2,0\rangle_{(c)}$.
    Note that Fig.~\ref{S_E_1,1}~B is identical to Fig.~\ref{S_E_20}~B and is reproduced here for completeness.
    Since the original plots were difficult to interpret, we produced an alternative visualization in which, for equal values of the $\int_D |\psi|^2dxdy$, the differences in the corresponding $\mu$ among the branches are plotted on the right for better clarity. The visualization and panels are similar to Fig.~\ref{mu2}.}
    \label{S_E_20}
\end{figure}
\begin{figure}
    \centering
    \includegraphics[width=0.5\linewidth]{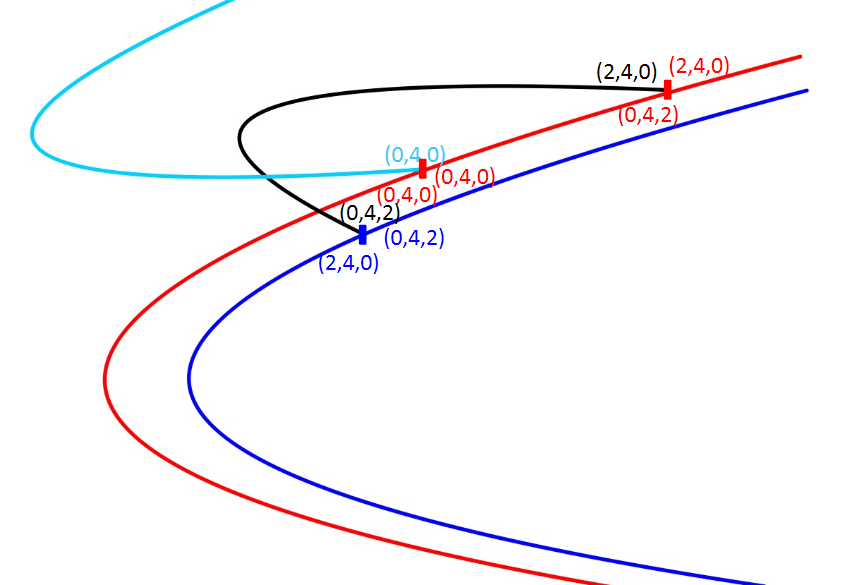}
    \caption{Bifurcation diagram in the $(\mu,\int_D |\psi|^2\,dx\,dy)$ plane with Fig.~\ref{S_E_20}. A simplified version is shown here for clearer visualization. Each tuple $(a,b,c)$ denotes, from left to right, the number of unstable eigenvalues (with purely imaginary $\omega$ from \S \ref{stability}) $a$, the number of eigenvalues exactly equal to zero $b$, and the number $c$ of eigenvalues that approach the origin from the stable side as the bifurcation point is approached. During bifurcations, these numbers may change, but their total sum $a+b+c$ remains constant before and after the bifurcation.}
    \label{fig:pla2}
\end{figure} 

\begin{itemize}
    \item \textbf{Branch from $|2,0\rangle_{(c)}$ (2-Dark Soliton Stripe)}
    
    The primary branch (Fig.~\ref{S_E_20}~A), originating from the $|2,0\rangle_{(c)}$ state, undergoes a series of bifurcations as $N=\int_D |\psi^2| dxdy$ increases. Fig.~\ref{S_E_20} B is identical to 
    Fig.~\ref{S_E_1,1} B, but is reproduced here for completeness in the discussion of the branches.
    
    \begin{itemize}
        \item \textbf{Bifurcation Sequence:}
        \begin{itemize}
            \item The first bifurcation occurs around $\mu \approx 0.377$, leading
            to the bifurcation of the rather non-canonical vortex quadrupole of Fig.~\ref{S_E_20}~D.
            \item The second bifurcation (Fig.~\ref{S_E_20}~C) takes place at $\mu \approx 0.442$. The resulting branch is observed to connect to a branch associated with the Fig.~\ref{S_E_1,1}~B$\equiv$ Fig.~\ref{S_E_20}~B state, as shown in Fig.~\ref{S_E_1,1}.
        \end{itemize}
    
        \item \textbf{Stability Analysis of the Primary Branch:}
        \begin{itemize}
            \item At the linear limit, the branch is stable with no unstable eigenvalues.
            The first bifurcation appears to again bear this flavor of a stable state leading to the
            bifurcation of another stable state without any eigenvalues crossing the origin; indeed,
            only 2 pairs of eigenvalues stay at the origin at this bifurcation point.
            After the second bifurcation, a  pair of unstable eigenvalues appears, rendering the
            2-stripe configuration unstable. While Fig.~\ref{S_E_20} contains the actual computed data, Fig.~\ref{fig:pla2} provides a qualitative representation to clarify the connectivity and stability transitions of the different branches.
            \item {\textbf{Remark:}} {
Tracking the stability of branch Fig.~\ref{S_E_20}~C reveals a significant evolution: while it possesses two exponentially unstable eigenvalues upon bifurcating from Fig.~\ref{S_E_20}~A, these eigenvalues are absent by the time it merges with Fig.~\ref{S_E_20}~B. To pinpoint when this transition occurs, we traced the evolution of the eigenvalues along the branch and confirmed that the loss of instability happens near at the turning point of Fig.~\ref{S_E_20}~C. This finding prompted a detailed investigation for additional solution branches that might originate from this critical point. While our investigation did not confirm the existence of any such branches, the underlying cause of the stabilization remains ambiguous: it could be an intrinsic feature of the turning point dynamics, such as a saddle-center bifurcation, or it might be influenced by a new branch that has so far eluded detection.
}
        \end{itemize}
    \end{itemize}

    \item \textbf{Branch from $|2,0\rangle_{(c)}+i|1,1\rangle_{(c)}$ }
    
    The primary branch (Fig.~\ref{sol_new}~A), originating from the 
$|2,0\rangle_{(c)} + i|1,1\rangle_{(c)}$ state and bearing a different quadrupole, undergoes a series of 
bifurcations as $N = \int_D |\psi|^2 \, dxdy$ increases. Here, we have not encountered
previously the branches Fig.~\ref{sol_new}~A and 
Fig.~\ref{sol_new}~E.
By contrast, the solutions in Fig.~\ref{sol_new}~B--D coincide with 
previously presented results 
(Fig.~\ref{S_E_1,1}~A $\equiv$ Fig.~\ref{sol_new}~B, 
Fig.~\ref{S_E_1,1}~B $\equiv$ Fig.~\ref{S_E_20}~B $\equiv$ Fig.~\ref{sol_new}~C, and
Fig.~\ref{S_E_20}~C $\equiv$ Fig.~\ref{sol_new}~D).

    \begin{itemize}
        \item \textbf{Bifurcation Sequence:}
        \begin{itemize}
            \item {The first bifurcation occurs around $\mu \approx 0.285$. The corresponding solution (Fig.~\ref{sol_new}~E) profile appears to interpolate between the two branches shown in Fig.~\ref{sol_new}~A and Fig.~\ref{sol_new}~D.}            
            \item At the second turning point of the branch Fig.~\ref{sol_new}~A, we note that the
           relevant  branch appears to regain its stability. As concerns this outcome (and, in general, the results at high $N =\int_D |\psi|^2 \, dxdy$ ) further investigation is required.
        \end{itemize}
    
        \item \textbf{Stability Analysis of the Primary Branch:}
        \begin{itemize}
            \item At the linear limit, the $|2,0\rangle_{(c)}+i|1,1\rangle_{(c)}$  branch is stable with no unstable eigenvalues. After the first bifurcation (resulting in the emergence of the waveform of Fig.~\ref{sol_new}~E), it develops a single pair of unstable eigenvalues. {We traced the dependence of the eigenvalues along the branch (Fig.~\ref{sol_new}~A) and found that, similar to the behavior in Fig.~\ref{S_E_20}~C, the instability-inducing
            real eigenvalue pair crosses the origin becoming imaginary at the final turning point. As in Fig.~\ref{S_E_20}~C, the most likely feature here is the presence of a saddle-center bifurcation at this turning point (although we have not been able to fully exclude the possibility of a pitchfork event).} While Fig.~\ref{sol_new} contains the actual computed data, Fig.~\ref{fig:pla3} provides a qualitative representation to clarify the connectivity and stability transitions of the different branches.
            \item\textbf{Remark:} {The solutions presented in Fig.~\ref{sol_new} B, C, and D are not new but correspond to previously reported results. Specifically, Fig.~\ref{sol_new}~B shows the same solution as Fig.~\ref{S_E_1,1}~A, Fig.~\ref{sol_new}~C corresponds to the solutions in Fig.~\ref{S_E_1,1}~B and Fig.~\ref{S_E_20}~B, and Fig.~\ref{sol_new}~D shows the same branch as Fig.~\ref{S_E_20}~C. Readers are directed to these referenced figures for detailed analyses.}
        \end{itemize}
    \end{itemize}
\end{itemize}

One can already detect the rapidly increasing complexity of the emerging bifurcation diagrams.
Accordingly, we do not proceed to further states here, although, in principle, the toolbox
developed herein can be directly adapted to such cases, as may be relevant in the future.

\begin{figure}[ht!]
    \centering
 \includegraphics[width=0.48\linewidth]{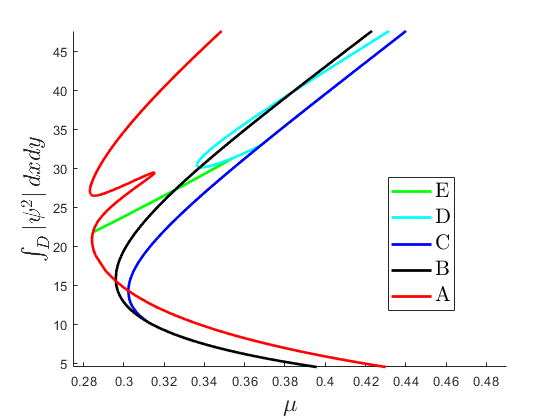}
  \includegraphics[width=0.48\linewidth]{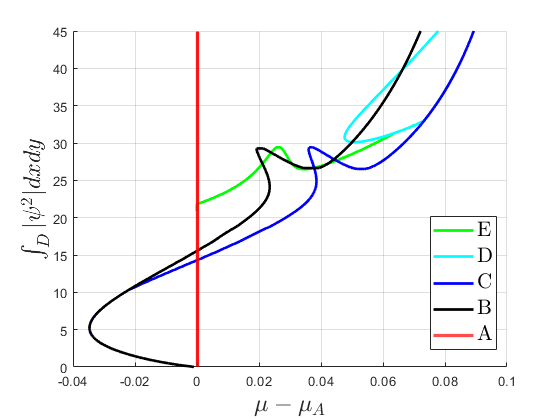}    \includegraphics[width=1\linewidth]{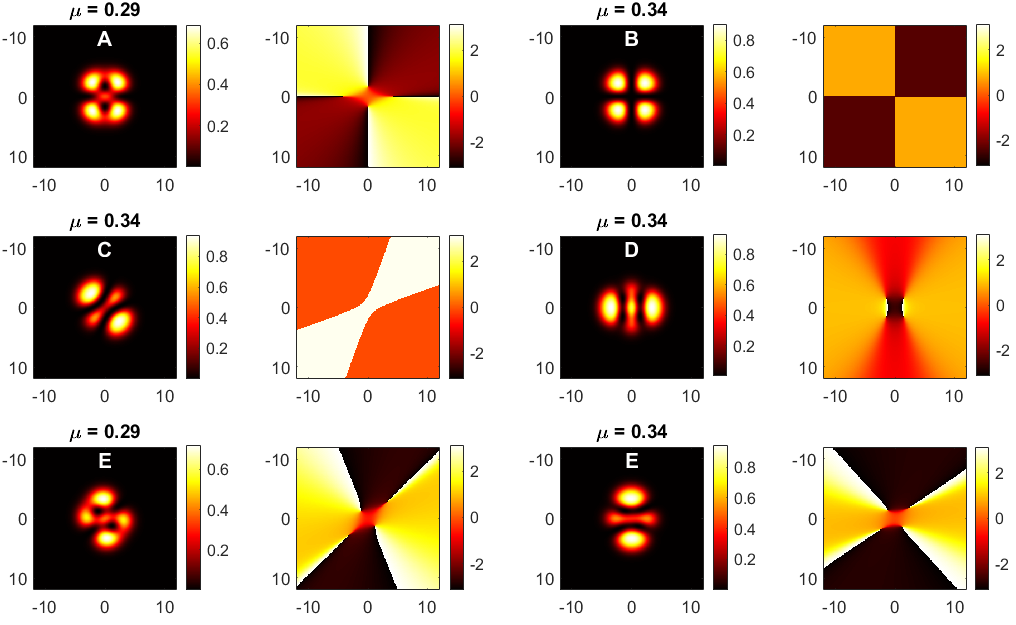}
    \caption{The first figure visualizes the bifurcations with respect to $\mu$, constructed for the branches connected with $|2,0\rangle_{(c)}+i|1,1\rangle_{(c)}$. Since the original plots were difficult to interpret, we produced an alternative visualization in which, for equal values of the $\int_D |\psi|^2dxdy$, the differences in the corresponding $\mu$ among the branches are plotted on the right for better clarity. The visualization and panels are similar to Fig.~\ref{mu2}. Note the equivalences: Fig.~\ref{S_E_1,1}~A $\equiv$ Fig.~\ref{sol_new}~B,  Fig.~\ref{S_E_1,1}~B $\equiv$ Fig.~\ref{S_E_20}~B $\equiv$ Fig.~\ref{sol_new}~C, and Fig.~\ref{S_E_20}~C $\equiv$ Fig.~\ref{sol_new}~D. These repeated panels are included for clarity and to emphasize the correspondence among branches across different continuation diagrams.}
    \label{sol_new}
\end{figure}

\begin{figure}[ht!]
    \centering
    \includegraphics[width=0.5\linewidth]{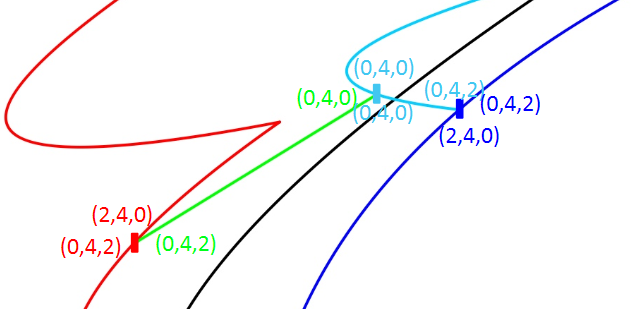}
    \caption{Bifurcation diagram in the $(\mu,\int_D |\psi|^2\,dx\,dy)$ plane with Fig.~\ref{sol_new}. A simplified version is shown here for clearer visualization. Each tuple $(a,b,c)$ denotes, from left to right, the number of unstable eigenvalues (with purely imaginary $\omega$ from \S \ref{stability}) $a$, the number of eigenvalues exactly equal to zero $b$, and the number $c$ of eigenvalues that approach the origin from the stable side as the bifurcation point is approached. During bifurcations, these numbers may change, but their total sum $a+b+c$ remains constant before and after the bifurcation.}
    \label{fig:pla3}
\end{figure}

\begin{table}[ht]
\centering
\scriptsize
\begin{tabular}{|p{5cm}|p{7.2cm}|p{4cm}|}
\hline
\textbf{Main Branch $\mu=2\Omega$} & \textbf{Bifurcations ($\mu$ values with Fig.)} & \textbf{Number of exponentially unstable eigenvalue pairs} \\
\hline

1-Dark Soliton Stripe ($|1,0\rangle_{(c)}$, Fig.~\ref{mu2}A) 
& $\mu \approx 0.171$ (Fig.~\ref{mu2}Middle), 
   0.41 (vortex dipole, Fig.~\ref{mu2}C), 
   0.655 (vortex tripole, Fig.~\ref{mu2}D) 
& 0 $\to$ 0 $\to$ 1  \\
\hline

Single-Charge Vortex ($|0,1\rangle_{(p)}$, Fig.~\ref{mu2}B) 
& $\mu \approx 0.078$ (Fig.~\ref{mu2}Middle) 
& 0 $\to$ 0 \\
\hline
\hline
\textbf{Main Branch $\mu=3\Omega$} & \textbf{Bifurcations ($\mu$ values with Fig.)} & \textbf{Number of exponentially unstable eigenvalue pairs} \\
\hline

Dark Soliton Cross ($|1,1\rangle_{(c)}$, Fig.~\ref{S_E_1,1}A) 
& $\mu \approx 0.312$ (Fig.~\ref{S_E_1,1}B), 
   0.38 (Fig.~\ref{S_E_1,1}Middle), 
   0.407 (Fig.~\ref{S_E_1,1}C), 
   0.619 (Fig.~\ref{S_E_1,1}D) 
& 1 $\to$ 0 $\to$ 0 $\to$ 1 $\to$ 2 \\
\hline

Charge-two Vortex ($|0,2\rangle_{(p)}$, Fig.~\ref{S_E_1,1}E) 
& $\mu \approx 0.296$ (Fig.~\ref{S_E_1,1}F), 
   0.275 (Fig.~\ref{S_E_1,1}Middle), 
   0.358 (Fig.~\ref{S_E_1,1}G) 
& 0 $\to$  0 $\to$ 0  $\to$ 0  \\
\hline

Ring Dark Soliton ($|1,0\rangle_{(p)}$, Fig.~\ref{S_E_10p}A) 
& $\mu \approx 0.349$ (Fig.~\ref{S_E_10p}B), 
   0.337 (merge, Fig.~\ref{S_E_10p}C), 
   0.363 (return, Fig.~\ref{S_E_10p}B), 
   0.52 (Fig.~\ref{S_E_10p}E), 
   0.79 (Fig.~\ref{S_E_10p}F) 
& 2 $\to$ 4 $\to$ 2 $\to$ 0 $\to$ 2 $\to$ 4 \\
\hline

Vortex Quadrupole ($|2,0\rangle_{(c)}+i|0,2\rangle_{(c)}$, Fig.~\ref{S_E_10p}C) 
& $\mu \approx 0.334$ (Fig.~\ref{S_E_10p}D), 
   0.337 (merge, Fig.~\ref{S_E_10p}A) 
& 0 $\to$ 2 $\to$ 2 \\
\hline

2-Dark Soliton Stripe ($|2,0\rangle_{(c)}$, Fig.~\ref{S_E_20}A) 
& $\mu \approx 0.377$ (Fig.~\ref{S_E_20}D), 
   0.442 (Fig.~\ref{S_E_20}C $\equiv$ Fig.~\ref{S_E_1,1}B) 
& 0 $\to$ 0 $\to$ 1 \\
\hline

$|2,0\rangle_{(c)}+i|1,1\rangle_{(c)}$ 
& $\mu \approx 0.285$ (Fig.~\ref{sol_new}D) 
& 0 $\to$ 1 $\to$ 0 \\
\hline

\end{tabular}
\caption{Compact summary of bifurcation sequences and number of unstable eigenvalue pairs for branches starts at $\mu=3\Omega=0.4$ and $\mu=3\Omega=0.6$.}\label{Table_mu3}
\end{table}

\section{Conclusions and Future Challenges}

In this work, we investigated a problem of significant current interest in the atomic physics community. Specifically, we considered the extended Gross-Pitaevskii (eGPE) model, which captures the interplay between mean-field effects and quantum fluctuations via the Lee-Huang-Yang correction. This interplay leads to the formation of quantum droplets, as well as a variety of other dynamical states, which have been observed to dynamically emerge and interact in a range of recent experiments~\cite{burchianti2020dual,cabrera2018quantum,cavicchioli2024dynamicalformationmultiplequantum,cheiney2018bright,d2019observation,semeghini2018self}.

Motivated by these developments, we analyzed the emerging families of solutions and their associated bifurcations within this model through the development of suitable numerical methods. Our methodology combined techniques tailored to either the 1D or 2D setting: the companion-based multi-level method and the homotopy grid expansion in 1D, and the dimension-by-dimension homotopy method in 2D. In the 1D case, these techniques produced quasi-1D solutions relevant to the 2D problem (i.e., solutions homogeneous in the transverse direction). In 2D, these approaches enabled the systematic identification of a broad class of solutions, most of which, to the best of our knowledge, are unprecedented in the context of the eGPE.  

While analogous solutions exist in the cubic GPE framework~\cite{charalampidis2018computing}—and are reproduced in the Appendix as a benchmark—the eGPE setting introduces additional complexities. In 1D, such complexities were previously understood to arise from the competition between attractive and repulsive nonlinearities, without giving rise to bifurcations. In contrast, our present work revealed a significantly richer bifurcation structure in 2D. Specifically, we observed instances of apparent pitchfork bifurcations where the associated branches remained either unstable or stable both before and after the bifurcation. We also detected pitchforks involving multiple eigenvalues approaching the origin concurrently at the bifurcation point. Moreover, supercritical pitchforks in the cubic case frequently transformed into subcritical pitchforks in the eGPE setting. Novel branch connections were also found (e.g., linking dark soliton stripes and vortex branches through intermediate ``middle'' branches), as well as bifurcations absent in the cubic case (e.g., splitting of a charge-2 vortex into two charge-1 vortices). Additionally, examples of saddle-center bifurcations were also discussed in the
context of this system. Lastly, these different bifurcation features
led branches (such as the ring dark solitons) that could never
be stable in the standard cubic GPE to be potentially stabilized
for a range of chemical potentials within the eGPE.

These findings, particularly at the bifurcation level, present several challenges for future numerical exploration. Quantifying and understanding the emerging bifurcation landscape in more detail remains an important direction. While the vast majority of instabilities were identified and corroborated, a small number of examples merit further clarification. These include subtle numerical effects associated with a rotational-invariance eigenpair (Fig.~\ref{mu2}~D), unexpected spectral behavior due to potential additional symmetries (Fig.~\ref{S_E_10p}~D), and changes in the number of unstable eigenvalues near fold points (Fig.~\ref{S_E_20}~C and Fig.~\ref{sol_new}~A). The latter suggests a possible bifurcation or an intrinsic turning-point phenomenon that remains to be fully clarified. These isolated spectral features warrant dedicated future analysis.  

While this work focused on some of the most fundamental branches, examining higher-order excited states is of considerable interest. Additionally, multiple extensions of the problem merit further study. {First, the original experiments were 3D in nature and employed the cubic-attractive/quartic-repulsive form of the 3D eGPE~\cite{cheiney2018bright}.} Thus, 3D generalizations of the present considerations, while computationally intensive, are both feasible and experimentally relevant~\cite{patrick3d}. In pursuing such extensions, recent studies have reported the modulational instability and discrete quantum droplets in quasi-one-dimensional optical lattices, 
obtained through reductions of the 3D extended Gross-Pitaevskii equation (eGPE)~\cite{otajonov2025modulational}, providing a useful starting point for further investigations. Second, we studied a reduced two-component model assuming similar spatial profiles for the components. Exploring the full two-component model, where components can differ, is an important future direction~\cite{simosstathis}. Performing corresponding multi-component computations in 2D and 3D, analogously to studies in the cubic nonlinearity case~\cite{patrick3d2c,CHARALAMPIDIS2020105255}, is also a natural avenue for future work. Lastly, while the above suggests that much remains to be
done at the existence, stability and overall bifurcation analysis
landscape, studies associated with
the dynamical evolution of such states, especially the unstable ones, 
in all of the above settings are also key in connection to acquiring
a detailed understanding of 
experimental observations. Such studies are currently in 
progress and will be reported in future publications.

\section*{Declarations}

\subsection*{Funding}
Wenrui Hao and Sun Lee were supported  by the National Science Foundation award DMS-2533995 and by the National Institutes of Health under award number 1R35GM146894. This research was also supported by the Huck Chair in AI Mathematical Modeling from Penn State University's Huck Institutes of the Life Sciences.

\subsection*{Conflict of Interest}
The authors declare that they have no conflict of interest.

\subsection*{Data Availability}
Data sharing is not applicable to this article, as no data sets were generated or analyzed.

\subsection*{Author contributions}
Conceptualization: Panayotis G. Kevrekidis, Wenrui Hao; 
Methodology: Sun Lee, Wenrui Hao; 
Formal analysis: Sun Lee, Panayotis G. Kevrekidis; 
Investigation, Visualization, and Writing – original draft: Sun Lee; 
Writing – review \& editing: Sun Lee, Panayotis G. Kevrekidis, Wenrui Hao; 
Supervision: Panayotis G. Kevrekidis, Wenrui Hao.

\subsection*{Acknowledgments}
The authors gratefully acknowledge the insightful discussions and valuable guidance provided by Prof.~Stathis Charalampidis (San Diego State University). 
This material is partially based upon work supported by the U.S. National Science Foundation under the award PHY-2408988 (PGK). This research was partly conducted while P.G.K. was 
visiting the Okinawa Institute of Science and
Technology (OIST) through the Theoretical Sciences Visiting Program (TSVP). 
This work was also supported by the Simons Foundation
[SFI-MPS-SFM-00011048, P.G.K.].

\newpage
\appendix
\section{Cubic Gross-Pitaevskii Benchmarks}
\label{sec:appendix_name}

\subsection{Gross–Pitaevskii equation in 1D}
\begin{figure}[ht]
    \centering
    \includegraphics[width=.45\textwidth]{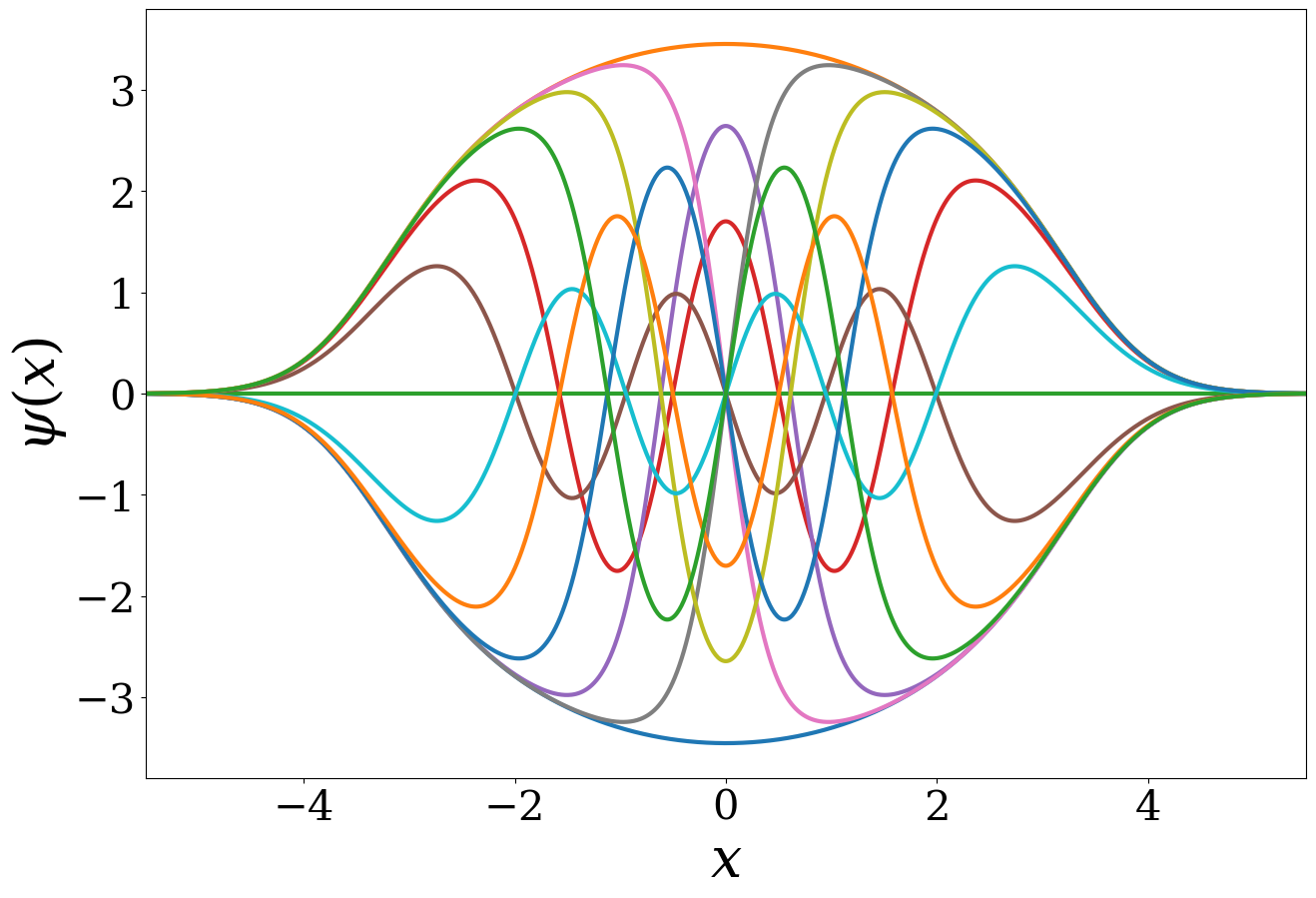}
    \includegraphics[width=.45\textwidth]{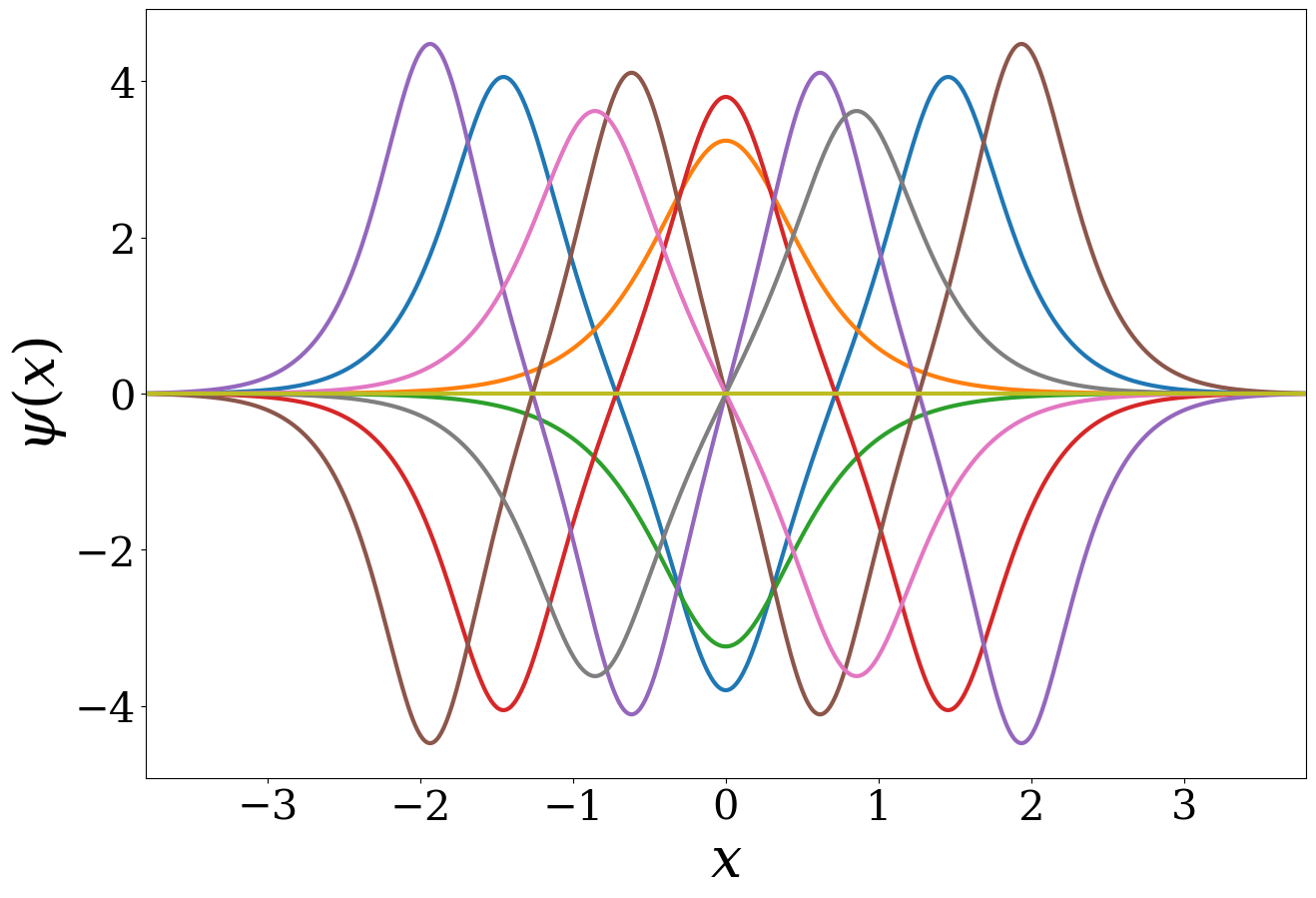}
    \caption{Numerical solutions of the one-dimensional Gross--Pitaevskii Eq.~\eqref{1dfin} are computed on a grid of 1025 points over domain $D$. \textbf{Left:} Repulsive case ($\sigma=-1$) with $\mu=12$ and $D=[-5.5,5.5]$, showing 13 solutions. \textbf{Right:} Attractive case ($\sigma=1$) with $\mu=-5$ and $D=[-3.8,3.8]$, showing 9 solutions.}
    \label{1dsols}
    
\end{figure}

\begin{figure}[ht]
    \centering
    \includegraphics[width=.4\textwidth]{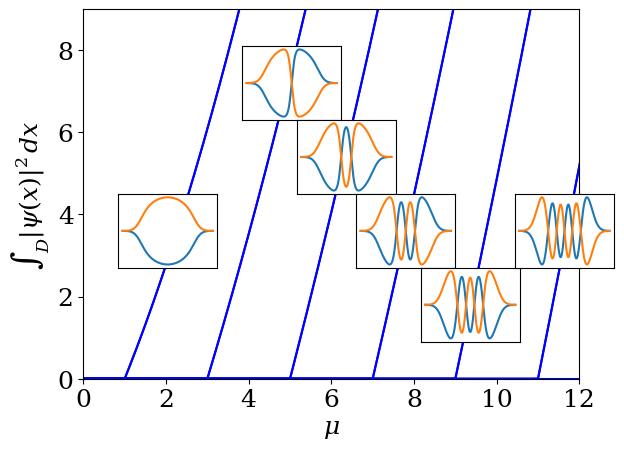}
    \includegraphics[width=.4\textwidth]{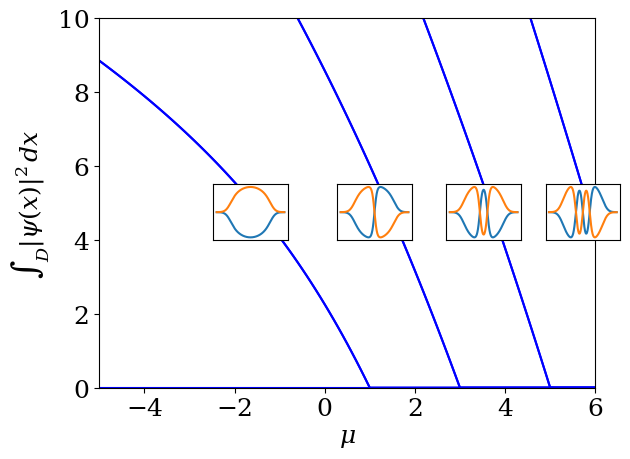}
    \caption{Bifurcation diagrams of $\int_{D} |\psi|^2 , dx$ versus $\mu$ for the one-dimensional Gross–Pitaevskii equation~\eqref{1dfin} in the repulsive case ($\sigma=-1$, left) and the attractive case ($\sigma=1$, right).}
    \label{example1}    
\end{figure}

In this section, we consider the one-dimensional Gross--Pitaevskii equation (GPE):
\begin{equation}
\label{NLS1d}
    i \,\partial_t \Psi(t,x) 
    \;=\; \,\Delta \Psi(t,x) \;-\; V(x)\,\Psi(t,x)+\; \sigma \,\bigl|\Psi(t,x)\bigr|^2 \,\Psi(t,x)
    \;, 
\quad (t,x)\in \mathbb{R}^+ \times (-d,d),
\end{equation}
with $\sigma \in \{+1, -1\}$ representing attractive and repulsive interactions, respectively. {We use the potential $V(x)=x^2$ 
and overall setup of Eq.~(\ref{NLS1d})
to match the scaling of ~\cite{alfimov2007nonlinear}, allowing direct reproduction and comparison of their results with our methods, which is a primary goal of this appendix.}
Following the standard ansatz (see, e.g., \cite{alfimov2007nonlinear})
\begin{equation}
\label{eq:ansatz}
    \Psi(t,x) \;=\; e^{i\mu \, t}\,\psi(x),
\end{equation}
we obtain the stationary GPE:
\begin{equation}
\label{1dfin}
    -\,\Delta \psi(x) \;+\; \bigl[V(x)-\mu\bigr]\,\psi(x) 
    \;-\; \sigma\,\bigl|\psi(x)\bigr|^2\,\psi(x) \;=\; 0.
\end{equation}
 Using the companion-based multi-level method detailed in \S \ref{cbm}, we obtained 9 solutions for the attractive case ($\sigma=1$) with $\mu=-5$, and 13 solutions for the repulsive case ($\sigma=-1$) with $\mu=12$. The solutions, on 1025 grid points, are illustrated in Fig. \ref{1dsols}. To facilitate a comparison with the findings of \cite{alfimov2007nonlinear}, we employed the homotopy continuation method to construct bifurcation diagrams of $N=\int |\Psi|^2 dx$ versus $\mu$, demonstrating consistency, as shown in Fig. \ref{example1}.

\subsection{Gross–Pitaevskii equation in 2D
repulsive case}
\begin{figure}
    \centering
    \includegraphics[width=.4\textwidth]{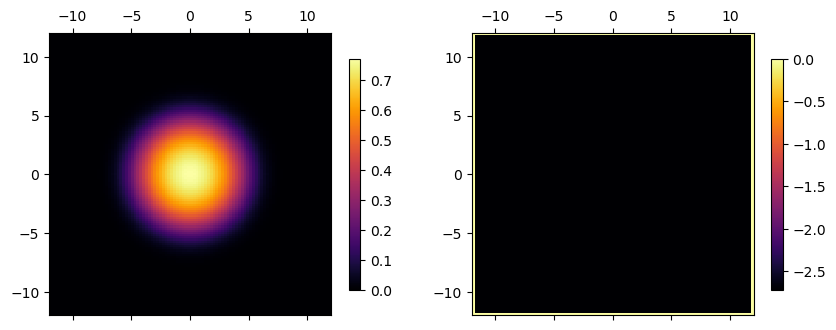}
    \includegraphics[width=.4\textwidth]{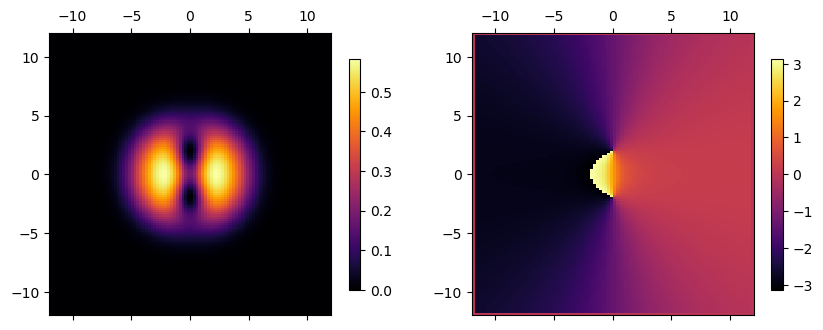}
    \includegraphics[width=.4\textwidth]{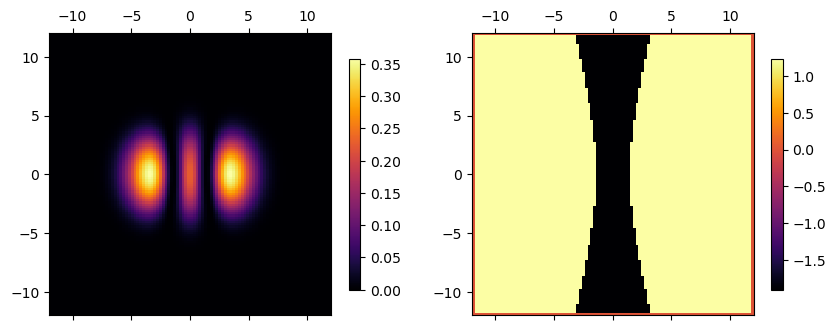}
    \includegraphics[width=.4\textwidth]{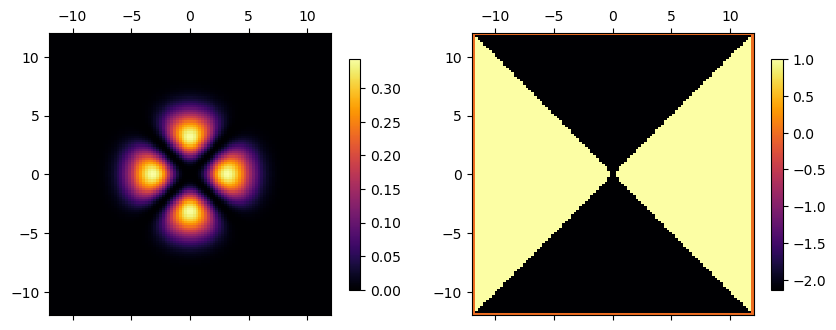}
    \includegraphics[width=.4\textwidth]{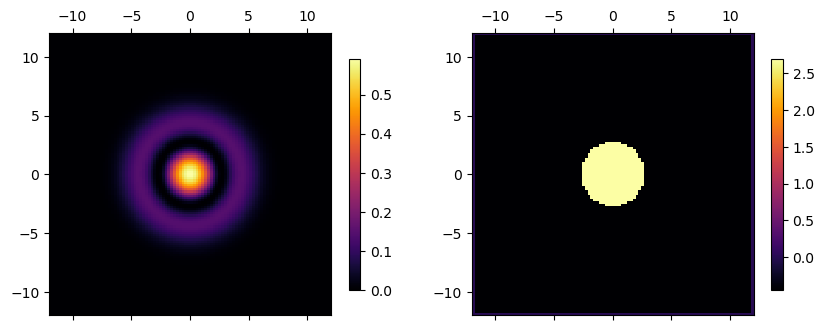}
    \includegraphics[width=.4\textwidth]{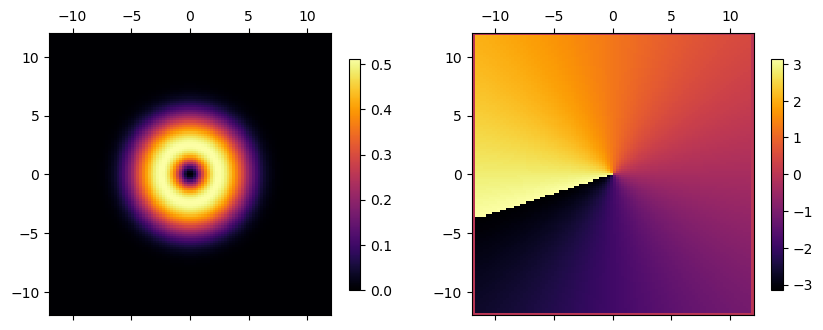}
    \includegraphics[width=.4\textwidth]{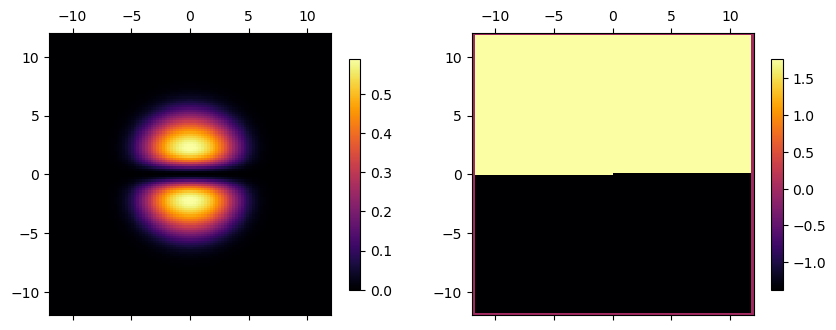}
    \includegraphics[width=.4\textwidth]{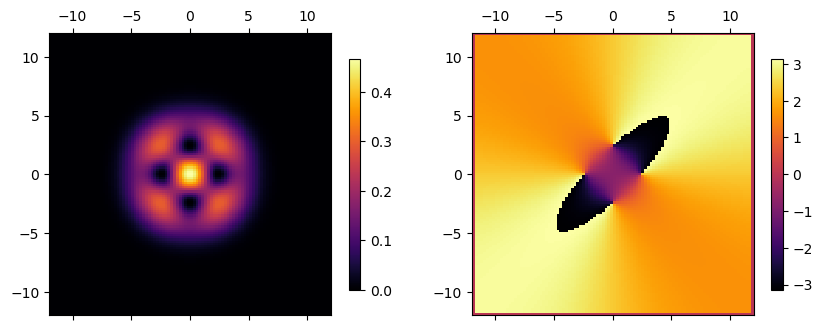}
    \includegraphics[width=.4\textwidth]{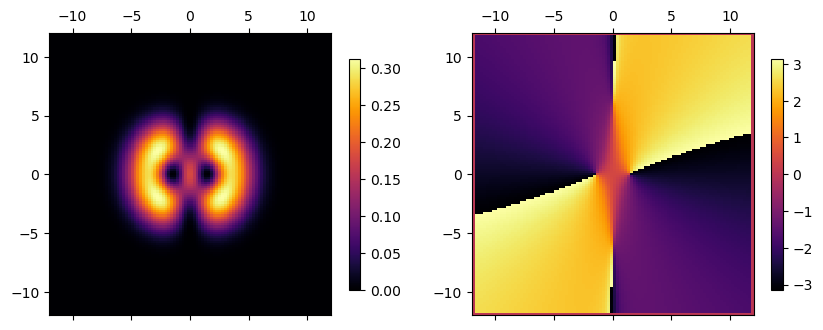}
    \includegraphics[width=.4\textwidth]{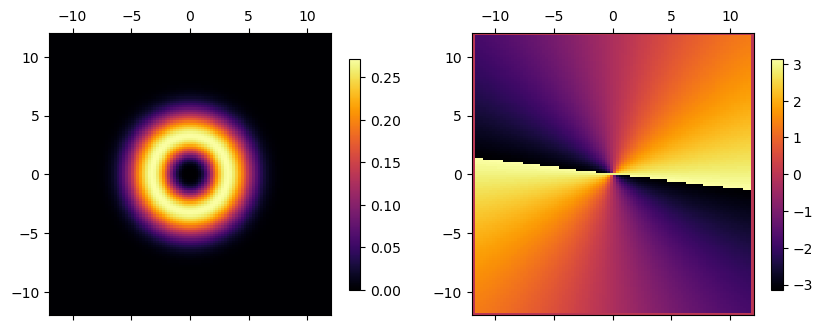}
    \caption{Numerical solutions of Gross–Pitaevskii equation in 2D. We choose $D=(-12,12)^2$, $\Omega =0.2$, and $\mu =0.8$. The grid points are $N_x=129$ and $N_y=129$. $|\psi|^2=\psi^2_\mathbb{R}+\psi^2_\mathbb{C}$ and the phase angle refers to the angle formed by $\theta=\tan^{-1}( \psi_\mathbb{C}/\psi_\mathbb{R})$.
    }
    \label{2dex1}
\end{figure}

The Gross–Pitaevskii equation in 2D is described  by
\begin{equation}
    i \partial_t \Psi =-\frac{\triangle \Psi }{2} + |\Psi|^2\Psi+V(r) \Psi , \quad \mbox{ in } D \times \mathbb{R}^+.
\end{equation}
{Here, $V(r)=\frac{\Omega^2 r^2}{2}$, with the trap strength $\Omega$, and $r^2=x^2+y^2$.} The corresponding stationary equation becomes:
\begin{equation}
     0=-\frac{\triangle \psi}{2} + |\psi|^2\psi + \frac{\Omega^2r^2}{2}\psi - \mu \psi 
\end{equation}
with Dirichlet boundary conditions. For our computations, we set $D=(-12,12)^2$, $\Omega =0.2$, and $\mu =0.8$. Employing the dimension-by-dimension homotopy method in Section \ref{homod} and the Homotopy Grid Expansion method in Section \ref{fin}, we obtained solutions in 2D, as depicted in Fig. \ref{2dex1}. These solutions are obtained on grid points $N_x=129$ and $N_y=129$. These results align with those reported in the literature, as shown in \cite{charalampidis2018computing}.

\section{Analysis of rotational symmetry and coordinate effects}\label{polar_coord}
We observed that when a solution $\psi$ is rotated around the origin, the squared magnitude $|\psi|^2$ may or may not exhibit rotational invariance. Specifically, in Fig.~\ref{mu2}~B, the quantity $|\psi|^2$ remains invariant under rotation, and consequently, the only zero eigenvalue pair corresponds to the phase mode. In contrast, for the case shown in Fig.~\ref{mu2}~A, where $|\psi|^2$ is not rotation-invariant (i.e., when a new ---rather than
the same--- solution is
produced through such a rotation), we identified an additional zero eigenvalue pair associated with the rotational mode.

\begin{figure}[ht!]
    \centering
    \includegraphics[width=0.7\linewidth]{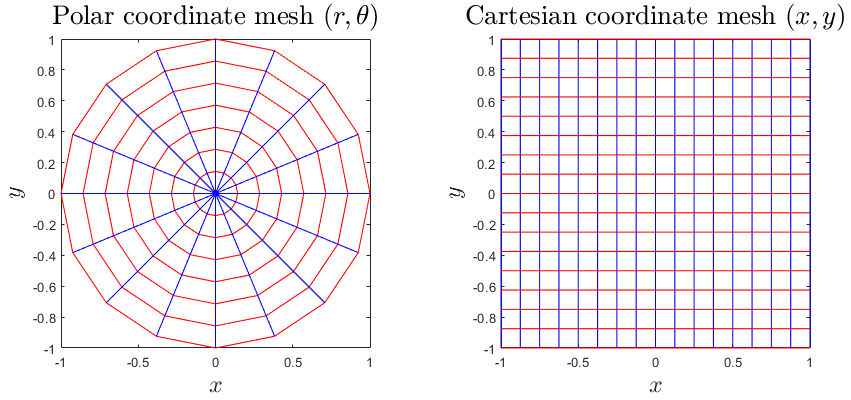}
    \caption{Visualization of Cartesian and polar meshes used to analyze rotational symmetry.}
    \label{fig:pol_car}
\end{figure}

Based on this observation, we tracked how the eigenvalues evolve as $\int_D |\psi|^2\,dx\,dy$ increases for the non–rotation-invariant case and found that the rotation-related eigenvalue tends to move away from zero as $\int_D |\psi|^2\,dx\,dy$ becomes larger. We initially suspected that this behavior was induced by the use of a Cartesian grid, which does not preserve rotational symmetry. To further examine the relationship between the solution structure and its spectral properties, we compared the results obtained on Cartesian and polar meshes, visualized in Fig.~\ref{fig:pol_car}. When the computation was repeated on a polar grid, the magnitude of the rotation-related eigenvalue decreased, thereby supporting our hypothesis.

However, even in the polar-coordinate formulation, as $\int_D |\psi|^2\,dx\,dy$ increases, the corresponding eigenvalue still gradually drifts away from zero. We attribute this residual effect to the nonuniform resolution of the polar grid: as the radius increases, the mesh becomes progressively coarser, and the cumulative discretization error near the outer boundary grows. This error accumulation likely causes the deviation of the eigenvalue from zero. In particular, as $\int_D |\psi|^2\,dx\,dy$ increases, the solution extends further toward the outer region of the computational domain, where the polar grid is coarser, leading to a gradual loss of rotational invariance.

As a potential direction for future work, we plan to construct an adaptive polar mesh that refines the grid resolution with increasing radius. Although the Laplacian computation and other numerical procedures will become more complex under this discretization, we plan to carefully design the scheme to maintain numerical stability and efficiency. Successfully overcoming these challenges may allow for a more accurate preservation of rotational symmetry and further reduce the artificial eigenvalue (weak) drift observed in the current analysis as $\int_D |\psi|^2\,dx\,dy$ becomes large.

\bibliographystyle{plain}
\bibliography{references}

\end{document}